\documentclass[twocolumn,trackchanges]{aastex701}
\usepackage[colorinlistoftodos]{todonotes}
\usepackage{amsmath}
\newcommand{\be}{\begin{eqnarray}}
\newcommand{\ee}{\end{eqnarray}}

\usepackage{CJKutf8}
\usepackage{hyperref}
\usepackage{xcolor}
\newcommand{\rich}[1]{\textcolor{black}{#1}}

\begin{document}
	
\title{Tides in Massive Binaries: Numerical Solutions and Semi-Analytical Comparisons}

\begin{CJK*}{UTF8}{gbsn}


\author[0000-0001-9037-6180,gname=Sun,sname=Meng]{Meng Sun(孙萌)}

\affiliation{National Astronomical Observatories, Chinese Academy of Sciences, 20A Datun Road, Chaoyang District, Beijing 100101, China}
\affiliation{Center for Interdisciplinary Exploration and Research in Astrophysics (CIERA), Northwestern University, 1800 Sherman Ave, Evanston, IL 60201, USA}
\email[show]{sunmeng@nao.cas.cn}

\author[0009-0002-2514-9584, gname=Hongbo, sname=Xia]{Hongbo Xia(夏宏博)}
\affiliation{Center for Interdisciplinary Exploration and Research in Astrophysics (CIERA), Northwestern University, 1800 Sherman Ave, Evanston, IL 60201, USA}
\email{hongboxia2027@u.northwestern.edu}

\author[0000-0001-6692-6410,gname=Gossage,sname=Seth]{Seth Gossage}
\affiliation{Center for Interdisciplinary Exploration and Research in Astrophysics (CIERA), Northwestern University, 1800 Sherman Ave, Evanston, IL 60201, USA}
\affiliation{NSF-Simons AI Institute for the Sky (SkAI), 172 E. Chestnut Street, Chicago, IL 60611, USA}
\email{seth.gossage@northwestern.edu}

\author[0000-0001-9236-5469,gname=Kalogera,sname=Vicky]{Vicky Kalogera}
\affiliation{Center for Interdisciplinary Exploration and Research in Astrophysics (CIERA), Northwestern University, 1800 Sherman Ave, Evanston, IL 60201, USA}
\affiliation{NSF-Simons AI Institute for the Sky (SkAI), 172 E. Chestnut Street, Chicago, IL 60611, USA}
\affiliation{Department of Physics \& Astronomy, Northwestern University, 2145 Sheridan Road, Evanston, IL 60208}
\email{vicky@northwestern.edu}

\author[0000-0002-2874-2706,gname=Jifeng,sname=Liu]{Jifeng Liu(刘继峰)}
\affiliation{National Astronomical Observatories, Chinese Academy of Sciences, 20A Datun Road, Chaoyang District, Beijing 100101, China}
\affiliation{School of Astronomy and Space Science, University of Chinese Academy of Sciences, Beijing 100049, China}
\affiliation{Institute for Frontiers in Astronomy and Astrophysics, Beijing Normal University, Beijing 102206, China}
\affiliation{New Cornerstone Science Laboratory, National Astronomical Observatories, Chinese Academy of Sciences, Beijing 100012, China}
\email{jfliu@nao.cas.cn}

\author[0000-0003-4474-6528,gname=Rocha,sname=Kyle]{Kyle Akira Rocha}
\affiliation{Center for Interdisciplinary Exploration and Research in Astrophysics (CIERA), Northwestern University, 1800 Sherman Ave, Evanston, IL 60201, USA}
\affiliation{NSF-Simons AI Institute for the Sky (SkAI), 172 E. Chestnut Street, Chicago, IL 60611, USA}
\affiliation{Department of Physics \& Astronomy, Northwestern University, 2145 Sheridan Road, Evanston, IL 60208}
\email{kylerocha2024@u.northwestern.edu}

\author[0000-0002-2522-8605]{R. H. D. Townsend}
\affiliation{Department of Astronomy, University of Wisconsin-Madison, 475 N Charter St, Madison, WI 53706, USA}
\email{townsend@astro.wisc.edu}

\author[0000-0002-7464-498X]{Emmanouil Zapartas}
\affiliation{Institute of Astrophysics, FORTH, N. Plastira 100,  Heraklion, 70013, Greece}
\email{ezapartas@ia.forth.gr}

\begin{abstract}
We present a systematic comparison between the tidal secular evolution timescales predicted by the direct numerical method and those given by the commonly used semi-analytic prescriptions implemented in 1-D hydrostatic binary evolution codes. Our study focuses on binary systems with intermediate- to high-mass primaries ($M_1 = 5$–$50\,M_\odot$), companion masses between $1.4\,M_\odot$ and $10\,M_\odot$, and orbital periods ranging from 0.5 to 50 days. Before mass transfer, both approaches predict synchronization and orbital decay timescales that agree within $\sim$2 orders of magnitude and typically exceed the stellar main sequence lifetime, implying negligible tidal impact on secular orbital evolution. However, the implied dissipation channels differ, and the differences become more pronounced once mass transfer begins. To test the theoretical predictions against observations, we apply both approaches to the well-characterized PSR~J0045--7319 system, which has an orbital decay timescale of 0.5 Myr. The numerical solution reveals strong resonances with internal gravity waves, bringing the predicted orbital period change rate close to the observed value. In contrast, the semi-analytic prescriptions predict orbital decay timescales longer than the Hubble time. These results suggest that for population studies, modestly calibrated parameterized equations may suffice, but for individual systems, reliable interpretation requires direct numerical approaches.
\end{abstract}

\keywords{\uat{Binary stars}{154} --- \uat{Close binary stars}{254} --- \uat{Interacting binary stars}{801} --- \uat{Stellar astronomy}{1583} --- \uat{Stellar evolution}{1599}--- \uat{Stellar oscillations}{1617} --- \uat{Stellar
populations}{1622} --- \uat{Tidal interaction}{1699} --- \uat{Tides}{1702} --- \uat{Astronomy software}{1855} --- \uat{Stellar evolutionary models}{2046}}

\section{Introduction \label{sec:intro}}
Tidal interactions play a critical role in the evolution of close binary and planetary systems, facilitating angular momentum exchange between stellar spins and orbits, and often serving as the gateway to mass transfer and compact object formation. The best possible modeling of tidal processes can be very important for understanding the demographics of compact-object binaries (e.g., \citealt{Arras2003,wei2016,Yu2017a,Yu2017b,Lau2022}) and of planetary architectures (e.g., \citealt{Terquem1998,Wu2005a,Wu2005b,Barker2009,Vick2019,Barker2020,Dewberry2022,Dewberry2023,Weinberg2024}). Yet, simulation of individual binaries and large-scale population synthesis, tides are typically modeled using simplified prescriptions that are computationally efficient but physically approximate. These simplifications may be rarely calibrated against detailed models or observations.

Tidal responses are commonly divided into two components: equilibrium tides, which describe the quasi-static deformation of a star under its companion's gravity, and dynamical tides, which involve time-dependent oscillations excited when the forcing frequency resonates with the star's internal eigenmodes. Following this classification, it is also essential to consider the mechanisms by which tidal energy is dissipated within the stellar interior. The dominant damping process depends on the type of tide: equilibrium tides are primarily damped by turbulent viscosity in convective zones (i.e., convective damping; see, e.g., \citealt{Goldreich1977,Zahn1989,Verbunt1995,Goodman1997,Penev2009ApJ,Penev2011a,Penev2011b,Duguid2020a,Duguid2020b,Vidal2020a,Vidal2020b,Barker2021}), whereas dynamical tides are subject to radiative damping in radiative regions and may also undergo nonlinear damping due to wave breaking or parametric instability, a resonance-driven transfer of energy from the parent tide to daughter modes (see, e.g., \citealt{Wu2001,Barker2010,Weinberg2012}). In magnetized stars or planets, tidal flows interacting with magnetic fields can also generate electric currents, which dissipate through Ohmic diffusion, an additional channel of tidal energy loss \citep[e.g.,][]{Laine2008,Lin2018,Wei2018,Wei2022}. These dissipation mechanisms ultimately determine the rate at which tidal interactions drive changes in orbital separation and spin–orbit synchronization in astrophysical systems.

In detailed stellar evolution codes such as \texttt{MESA}, when tides are enabled the user can select among different prescriptions depending on the science case. The standard options are \texttt{Hut\_rad}, which assumes the whole star behaves radiatively and applies a global tidal timescale, and \texttt{Hut\_conv}, which follows \citet{Rasio1996,Hurley2002} in computing the equilibrium-tide contribution to the secular evolution timescale in the convective envelope. Thus, whether the system is governed by radiative damping or convective damping could be left as a user choice. Rapid population-synthesis frameworks  adopt prescriptions similar to those in \texttt{MESA}. In contrast, the version implemented in the next generation population synthesis framework \texttt{POSYDON} \citep{Fragos2023,Andrews2024} relies on a more physical \texttt{MESA} treatment of tides, where both radiative and convective prescriptions are computed simultaneously and the shorter timescale is adopted based on the stellar structure rather than on a user-specified option. We hereafter refer to these implementations as semi-analytical {prescriptions}.

In this work, we compare the secular evolution timescales predicted by semi-analytical prescription with that computed using the \texttt{GYRE-tides} code \citep{Sun2023}, \rich{an extension to the \texttt{GYRE} stellar oscillation code \citep{Townsend2013,Townsend2018,Goldstein2020}} that numerically solves the inhomogeneous, nonadiabatic, forced oscillation equations for stellar tidal responses. We apply both prescriptions to a suite of evolving binary models spanning a range of stellar masses and orbital separations, and quantify the tidal synchronization and semimajor-axis evolution timescales. Our goal is to systematically assess the secular impact of different tidal prescriptions across a broad parameter space. In doing so, we aim to determine whether current semi-analytical treatments reliably calculate the efficiency of tidal dissipation, or whether revisions are warranted. Because direct numerical methods are computationally too expensive to be applied across full binary grids, our study provides a benchmark to guide the development of more accurate yet tractable prescriptions. In the long term, such benchmarks could also motivate surrogate approaches, for instance machine-learning emulators trained on numerical calculations, to bridge the gap between accuracy and efficiency.

This paper is organized as follows. In Section~\ref{sec:methods}, we summarize the stellar models adopted in this study and provide an overview of the tidal prescriptions, including a brief historical perspective on their development. Section~\ref{sec:results} presents a comparison between a semi-analytic tidal prescription implemented in binary population synthesis models and numerical method across a set of representative binary evolutionary tracks. In Section~\ref{sec:discussions}, we validate the numerical results against the observed orbital decay of the PSR~J0045--7319 system, a binary consisting of a B-type main-sequence star and a radio pulsar. We then discuss, in a more general context, the possibility of resonance locking prior to Roche-lobe overflow (RLOF). We conclude in Section~\ref{sec:conclusions}.

\section{Methods\label{sec:methods}}
\subsection{Stellar Evolution Modeling with MESA}

For our study, we adopt the \texttt{MESA} binary evolution module (version 11701; \citealt{Paxton2011,Paxton2013,Paxton2015,Paxton2018,Paxton2019,Jermyn2023}) with solar metallicity $Z = 0.0142$ \citep{Asplund2009} and $Y = 0.2703$. These settings are integrated within the modified \texttt{POSYDON} version~2 framework to simulate the evolution of selected massive binary systems. While \texttt{POSYDON} is typically used for population synthesis studies, we instead utilize its built-in \texttt{MESA} inlists to evolve a set of benchmark binaries, including systems with initial configurations such as $M_1 = 5\,M_\odot$, $M_2 = 1.4\,M_\odot$, and more massive combinations up to $M_1 = 50\,M_\odot$. Throughout this paper, we use subscripts ``1” and ``2” to denote the initially more massive and less massive stars, respectively. These simulations are initialized with a range of orbital periods $P_{\rm orb} = 0.5 - 50$ days. We choose to build upon the \texttt{POSYDON-MESA} inlists because they incorporate a comprehensive treatment of the micro- and macrophysics (e.g. equation of state, opacity table, nuclear reaction network) relevant to massive stars, including stellar winds, convection, rotation, and various mixing processes. The following paragraphs revisit several of these key physical ingredients.

Stellar winds are implemented following a hybrid prescription that varies with effective temperature $T_{\mathrm{eff}}$ and evolutionary phase. For stars with lower initial masses $M_{\mathrm{initial}} < 8\,M_\odot$ and $T_{\mathrm{eff}} < 8000\,\mathrm{K}$, during the RGB phase, the Reimers wind \citep{Reimers1975} is used with a fixed scaling parameter, while during the AGB phase, the Blöcker wind \citep{Bloecker1995} is adopted. For $T_{\mathrm{eff}} > 12000$ K, the \texttt{MESA Dutch} prescription is applied. For intermediate temperature $8000\,\mathrm{K}<T_{\mathrm{eff}}<12000\,\mathrm{K}$, the wind rates are linearly interpolated to ensure smooth transitions. For massive stars with $M_{\mathrm{initial}} \geq 8\,M_\odot$, we adopt the \texttt{Dutch} wind prescription, which combines the formulae of \citealt{deJager1988,Nugis2000,Vink2001}, depending on the effective temperature and surface composition.

To prevent stars from reaching critical rotation, \texttt{MESA} adopts a boosted wind prescription: when a star approaches its critical surface rotation rate, its stellar wind is artificially enhanced, effectively removing excess angular momentum. During mass transfer, this mechanism ensures that the accretor maintains a sub-critical rotation rate, but it also significantly limits the amount of mass the accretor can retain. As a result, the overall mass transfer process becomes highly non-conservative, with the majority of the transferred mass being lost from the system (see \citealt{Sun2024a,Sun2024b,Rocha2024,Zapartas2025}).

Stellar models adopt mixing-length theory (MLT) with a solar-calibrated parameter $\alpha_{\rm MLT} = 1.93$. In radiation-dominated superadiabatic zones, \texttt{MLT++} \citep{Paxton2013} is used to improve convergence. Convective boundaries are determined using the Ledoux criterion. Rotation, {rotational mixing} and angular momentum transport via Tayler-Spruit dynamo \citep{Spruit2002} are considered in making the star models. The convective overshoot is described in the exponential-decay formalism \citep{Herwig2000}. The overshoot parameter $f_{\rm ov}$ adopted is mass-dependent: 0.016 for zero-age main-sequence (ZAMS) mass $M_{\rm initial} < 4\,M_\odot$ and 0.0415 for $M_{\rm initial} > 8\,M_\odot$, with linear interpolation in between. No additional semiconvective mixing is included beyond what is already captured by convective premixing.

\subsection{Tidal Dissipation Models}

Physically, tidal synchronization and the evolution of the orbital semimajor axis are two aspects of the same process. For example, consider a non-rotating star orbited by a companion: gravitational tides act to synchronize the star's spin with the orbital motion, drawing angular momentum from the orbit and transferring it to the star, thereby shrinking the orbit. Conversely, if the star initially rotates faster than the orbital motion of its companion, the same tidal torque extracts spin angular momentum and deposits it into the orbit, causing the orbit to expand.

Informed by the early work of \citet{Darwin1880}, who first linked tidal dissipation to orbital angular momentum exchange in viscous bodies, several approaches have been developed to quantify stellar tidal response and its long-term impact on orbital evolution. These methods differ in their treatment of the underlying physics, particularly whether they adopt an averaged, parameterized dissipation model or attempt to numerically calculate the response.

One of the earliest and most influential formulations of equilibrium tide theory was introduced by \citet{Goldreich1966}, who proposed the use of a constant tidal quality factor $Q$ to quantify the fraction of tidal energy dissipated per orbital cycle. In this model, the tidal phase lag is assumed to be frequency-independent, resulting in a fixed ratio between energy loss and energy stored in the tidal deformation. The simplicity and intuitive appeal of this approach have made it a cornerstone of tidal modeling in both planetary and stellar contexts.

Another closely related formulation was later developed by \citet{Alexander1973}, who introduced a constant time lag $\tau$ model. In this framework, the tidal bulge lags behind the line connecting the two bodies by a fixed time interval, rather than a fixed angle, which naturally leads to a phase lag that scales linearly with the tidal forcing frequency. The two models are connected by the relation $Q^{-1} \approx 2\omega_\mathrm{f} \tau$ in the low-frequency limit and the equilibrium tide assumption, where $\omega_\mathrm{f}$ is the tidal forcing frequency, but they diverge in their scaling and physical interpretation at higher frequencies.

The constant time lag model was further extended by \citet{Hut1981}, who derived closed-form secular evolution equations for $({\rm d}a/{\rm dt})_{\rm sec}$, $({\rm d}e/{\rm dt})_{\rm sec}$, and $({\rm d}\Omega_{\rm rot}/{\rm dt})_{\rm sec}$ (representing the orbital averaged secular rates of change of the semimajor axis $a$, eccentricity $e$, and stellar rotation rate, respectively) valid at arbitrary eccentricities and rotation rates $\Omega_{\rm rot}$. This weak friction approximation remains one of the most widely used tidal prescriptions in population synthesis studies, where it is typically implemented in the form of Hut’s analytic expressions. While this framework provides a convenient estimate of spin–orbit coupling efficiency, its assumption of constant time lag cannot capture frequency-dependent dissipation, and may significantly underestimate tidal effects in systems where dynamical tides are dominant. Modern implementations (e.g., \citealt{Fragos2023}) go beyond Hut’s equilibrium-tide formulation with viscous damping alone. They examine the stellar structure at each timestep and apply different prescriptions depending on the region: equilibrium tides damped by turbulent viscosity in convective zones, and dynamical tides damped by radiative diffusion in radiative zones.

This resonant picture of dynamical tides has its origins in the earlier theoretical framework of \citet{Cowling1941}, who investigated the non-radial oscillations of polytropic stars and identified the possibility of resonance between orbital forcing and internal modes. Later, \citet{Zahn1970, Zahn1975, Zahn1977} modeled the resonant excitation of internal gravity waves in stars with radiative envelopes. These waves are excited by the time-dependent tidal potential. When the forcing frequency of the tide is close to the frequency spectrum of g-modes in the star, resonant excitation occurs. The excited modes then dissipate their energy mainly through radiative diffusion, giving rise to frequency-dependent tidal torques. In this work, we use the term \textit{internal gravity wave} to describe tidally excited waves in the radiative zone. When damping is weak, such waves correspond to standing g-modes. However, in the strong-damping regime they may not form global standing modes, and instead behave as traveling waves. 

In particular, \citet{Zahn1975,Zahn1977} introduced the tidal torque constant $E_2$ to quantify the coupling strength between the tidal potential and the g-mode spectrum in the stellar radiative zone. Later works often interpolated or extrapolated $E_2$ values in regimes where this was physically unjustified, sometimes introducing order-of-magnitude errors in torque estimates \citep[see comments in][]{Kushnir2017}.

\citet{Savonije1983, Savonije1984} extended the formalism by numerically solving the nonadiabatic forced oscillation equations for massive stars at various evolutionary stages. Their results confirmed that tidal dissipation is generally weak unless the system encounters a resonance, which can substantially accelerate orbital circularization and spin synchronization, particularly in evolved stars where structural changes (e.g., the growth of semi-convective regions) can enhance mode coupling. A related isentropic treatment of forced oscillations was presented by \citet{Polfliet1990}.

To describe the strong frequency dependence that arises near resonances, \citet{Smeyers1998}, \citet{Willems2003}, \citet{Willems2010} and \citet{Valsecchi2013} applied a two-time variable expansion method to capture the secular evolution induced by individual resonant g-modes. In this mode-based framework, the tidal potential is decomposed into its Fourier components, and for each forcing frequency, a corresponding set of linear forced oscillation equations is numerically solved to compute the amplitude and phase of the stellar response. These solutions together characterize the total tidal response. The resulting mode properties are then inserted into secular evolution equations to determine the long-term changes in orbital parameters.

An alternative semi-analytic method was developed by \citet{Goodman1998}, who solved the linearized tidal response equations under the Cowling approximation. By evaluating mode amplitudes and damping rates, the method yields a detailed dissipation spectrum $\dot{E}_{\mathrm{tide}}(\Omega_\mathrm{orb})$ (where $\Omega_\mathrm{orb}$ is the mean orbital angular frequency) that captures both resonant and off-resonant behavior. This approach provides valuable insight into how structural properties, specifically at the radiative–convective boundary (RCB), determine the efficiency of tidal dissipation for solar-like stars.

There are two main classes of mode-based numerical methods for solving the linear tidal response. In the direct solution approach, the star’s response is evaluated by solving the forced oscillation equations as a two-point boundary value problem (BVP), with the tidal potential treated as an inhomogeneous term. This method directly computes the global response of the star at each forcing frequency. Inspired by \citet{Valsecchi2013}, a representative implementation of this approach is the \texttt{GYRE-tides} code introduced by \citet{Sun2023}, which computes nonadiabatic solutions using a BVP solver based on multiple shooting.

Alternatively, in the modal decomposition approach, the tidal response is expressed as a linear superposition of the star’s adiabatic eigenmodes. The amplitude of each mode is determined by evaluating an overlap integral that describes the spatial coupling between the tidal potential and the eigenfunction, a Hansen coefficient that characterizes the temporal coupling between the tidal potential and the orbital harmonics, and a Lorentzian factor that quantifies the temporal resonance between a given harmonic and a given mode \citep{Press1977,Burkart2012}. As highlighted by \citet{Townsend2023}, the two approaches can yield inconsistent results for the global torque exerted on the star, especially near the stellar surface where nonadiabatic effects are strong or the forcing is far from resonance. These discrepancies mainly reflect the fact that the eigenfunctions of the linear adiabatic free oscillation equations are not, in general, eigenfunctions of the corresponding non-adiabatic problem. As a result, a modal decomposition using adiabatic eigenmodes can yield systematically different stellar torques compared to a direct solution of the non-adiabatic forced oscillation equations (see \citealt{Dewberry2024} for a detailed discussion). The direct solution method is generally more robust, whereas the modal decomposition approach may suffer from incomplete surface responses. \citet{Townsend2023} proposed a correction to the Lorentzian profile by more accurately treating the nonadiabatic contributions to the modal damping rate, in order to reduce the discrepancy between the two numerical methods.

However, such mode-based numerical methods are computationally expensive and highly sensitive to the details of the stellar structure \citep{Townsend2025}, making it difficult to derive general-purpose fitting functions near resonances, where the tidal response can vary rapidly. Most population synthesis studies treat tidal interactions by fitting $E_2$. This parameter approximates the torque exerted on stars by radiative damping of dynamical tides. The tidal evolution equations of \citet{Hut1981} were originally formulated for equilibrium tides with viscous damping. In later work, these equations are often adapted by replacing or adding dynamical tides as the main dissipation mechanism. In the next section, we summarize the equations used to compute secular evolution timescales, both from structure-averaged approaches and from direct numerical solutions.

\subsection{Revisiting Tidal Prescriptions in Population-Synthesis and Binary Evolution Codes}
\label{subsec:method:POSYDON-MESA}

In our \texttt{MESA} tracks, the code computes a synchronization timescale resulting from dynamical tides $t_{\rm dyn}$ with radiative damping. If the model has a convective zone, an additional timescale resulting from equilibrium tides $t_{\rm eq}$ with convective damping is computed. The shorter of the two is adopted as the global synchronization timescale $t_{\rm sync}$ and is used to update the stellar spin. This is different from the default \texttt{MESA} tidal prescription, which requires the user to specify in the inlist which dissipation mechanism should be applied. To make the treatment more physical, our implementation lets the code automatically determine which mechanism is effective in modifying the orbit. Mathematically, we define the synchronization timescale as

\begin{equation}
\begin{aligned}
\frac{1}{t_{\rm sync}} & = \max\left( \frac{1}{t_{\rm eq}},\ \frac{1}{t_{\rm dyn}}\right) \\
&= 3 \,\left(\frac{k}{T}\right)\, q^2\,\frac{MR^2}{I}\left(\frac{R}{a}\right)^6.
\label{eq:t_sync}
\end{aligned}
\end{equation}

Each synchronization timescale is inversely proportional to $k/T$, where $k$ is the dimensionless apsidal motion constant and $T$ is the characteristic tidal timescale, defined as the ratio of the square of the dynamical timescale to the tidal lag time. The binary mass ratio is $q=M_{\rm comp}/M$, while $M$, $R$, and $I$ denote the mass, radius, and moment of inertia of the star experiencing the tidal torque (e.g., $M_1$, $R_1$, $I_1$, together with $k_1$ and $T_1$, when tides are raised on star 1).

In the case of $e=0$, combining the $k/T$ term with the semimajor axis evolution formula of \citet{Hut1981}, the characteristic timescale for semimajor axis change can be written as
\begin{equation}
    \frac{1}{t_a} \;=\; 6 \,\left(\frac{k}{T}\right)\, q \,(1+q)\,\left(\frac{R}{a}\right)^8 \left(\frac{P_{\rm orb}}{P_{\rm rot}} - 1\right),
\label{eq:ta}
\end{equation}
where $P_{\rm rot}=2\pi/\Omega_\mathrm{rot}$ is the star rotation period.

\subsubsection{Dynamical tide with $E_2$ Fitting}

In the case of radiative damping acting on dynamical tides, the value of $k/T$ is proportional to $E_2$. Following the prescription from \citet{Qin2018} for hydrogen-rich stars, 
$E_2$ is highly sensitive to the ratio between the location of the radiative–convective boundary (RCB; i.e., the base of the radiative zone in massive stars, $r_{\rm RCB}$) and $R$. The synchronization timescale due to radiative damping is from:

\begin{equation}
\frac{1}{t_{\rm dyn}} \propto \left( \frac{k}{T} \right)_{\rm rad} \propto E_2 = 10^{-0.42} \left( {r_{\rm RCB}}/{R} \right)^{7.5}.
\end{equation}

\subsubsection{Equilibrium Tide with Convection-Zone Structure}

For convective damping acting on equilibrium tides, \texttt{MESA} follows the prescription of \citet{Rasio1996,Hurley2002} to compute the synchronization timescale: 
\begin{equation}
\frac{1}{t_{\rm eq}} \propto \left( \frac{k}{T} \right)_{\rm conv} = \frac{2}{21} \cdot \frac{f_{\rm conv}}{\tau_{\rm conv}} \cdot \frac{M_{\rm env}}{M}.
\label{equ:t_eq}
\end{equation}
Here, $\tau_{\rm conv}$ is the convective eddy turnover time. It is computed from the convective envelope structure itself, its mass and the positions of its upper and lower boundaries, via Eq.~(31) in \citet{Hurley2002}. 
Thus, the convective damping is explicitly structure dependent. $M_{\rm env}$ is the mass of the convection zone. Our modified \texttt{MESA} script evaluates this expression for all convective layers outside the stellar core, selecting the most efficient one (shortest $t_{\rm eq}$). The factor $f_{\rm conv} = \min\left(1, \left( \frac{P_{\rm tide}}{2 \tau_{\rm conv}} \right)^{\! b} \right)$ accounts for the frequency-dependent reduction of viscous damping; following the prescription of \citet{Zahn1989}, the exponent $b$ is typically set to 1 (\texttt{tidal\_reduction = 1.0} in our inlist). The tidal forcing period \( P_{\rm tide} \) is given by:
\begin{equation}
P_{\rm tide} = \left| \frac{1}{P_{\rm orb}} - \frac{1}{P_{\rm rot}} \right|^{-1}.
\end{equation}
Note that this prescription does not account for orbital harmonics or the azimuthal number, treating the tidal forcing as a single dominant frequency.

{Unlike \texttt{MESA} default tidal treatment, to cover all evolutionary phases and states, we compute equilibrium tide timescales layer by layer, evaluating up to the first $\sim 100$ dynamically significant convective layers exterior to the convective core. In this treatment, the eddy turnover time is set to zero in radiative zones and is radius dependent within convective regions, with values taken locally from the stellar structure.} Equilibrium tidal dissipation is typically strongest in the outermost convective envelope, where the synchronization timescale is shortest. Thin near-surface convective zones can appear in our stellar models, some of which may reflect genuine physical features (e.g., associated with iron-group opacity peaks), while others may be numerical artifacts. To avoid overestimating tidal efficiency due to these potentially artificial convection zones, we restrict our calculations of $t_{\rm eq}$ to convective regions that extend over at least 10 consecutive mesh zones following \citet{Fragos2023}.

\subsection{Tidal Treatment in \texttt{GYRE-tides}}
\label{subsec:method:gyre}

In this work, the direct-solution implementation within the \texttt{GYRE-tides} module is used to compute the stellar response to tidal forcing in binaries. This approach directly solves the inhomogeneous linear nonadiabatic oscillation equations, in which both the equilibrium tide and resonantly excited dynamical tide naturally emerge as parts of the solution, without requiring any decomposition into normal modes. Both radiative damping and convective damping are included; in the latter case, the eddy turnover time is reduced by a linear-in-period factor following the Equation (68) of \citet{Willems2010} and {in our implementation, the viscosity term $\nu_{\rm GYRE}$ is radius dependent}:
\begin{equation}
\nu_{\rm GYRE} = \frac{L^2}{\tau_{\rm conv}} \left[1 + \left( \frac{\tau_{\rm conv} \sigma_{m,k}}{2\pi} \right)^s \right]^{-1},
\end{equation}
where $L$ is the {local} mixing length, $\sigma_{m,k} = k\Omega_{\rm orb} - m\Omega_{\rm rot}$ is the tidal forcing frequency, and $s$ is typically set to $1$ (a slightly different approach than Zahn's prescription). $m$ is the azimuthal wavenumber, and $k$ is the Fourier index associated with the tidal forcing frequency decomposition (i.e. orbital harmonics). {We note that the parameters $b$ and $s$ remain uncertain and debated: stronger high-frequency suppression has been proposed in classical works \citep[e.g.,][]{Goldreich1977,Goodman1997}, whereas more recent studies suggest that a different reduction may be necessary \citep[e.g.,][]{Terquem2021,Barker2021,Terquem2023}.}

The tidal potential from a companion is expanded into spherical harmonics and orbital harmonics indexed by $(\ell, m, k)$. Each partial tidal potential $\Phi_{\mathrm{T};\,\ell,m,k}$ gives rise to a set of six inhomogeneous differential equations (see Appendix~E of \citealt{Sun2023}) that describe the radial and horizontal displacements of the tidally driven wave, as well as the perturbations to the thermal physical quantities and the gravitational potential. In the DS framework, the differential equations themselves retain almost the same structure as in the free-oscillation case; the tidal potential enters only through an inhomogeneous outer boundary condition (see Equation (15) of \citealt{Sun2023}). These are solved using a multiple shooting algorithm in the \texttt{GYRE} framework, with $\sigma_{m,k}$ supplied as an input parameter. 

We assume the star is spherically symmetric and neglect both the Coriolis and centrifugal forces, restricting the current implementation to aligned systems without spin–orbit misalignment. However, uniform rotation is included through the definition of the forcing frequency and in computing the stellar response. The perturbation amplitudes scale linearly with the dimensionless tidal forcing strength $\varepsilon_\mathrm{T}=(R/a)^3q$ (Equation 10 of \citealt{Sun2023}). This linearity allows the response to be rescaled for binary systems with different $q$ and $a$, as long as the forcing frequency and stellar model remain unchanged.

For each forcing term, \texttt{GYRE-tides} returns the full numerical solution to the forced oscillation problem, including both the interior structure and surface perturbations such as the radial displacement $\xi_r$ and the Lagrangian radial flux perturbation $\delta F_{\rm rad}$. A key intermediate output is the complex tidal response function $F_{\ell,m,k}$. It is defined in Equation (50) of \citet{Willems2010}, with further details given in Equation (57) of \citet{Valsecchi2013}. This function encapsulates both the amplitude and the phase lag of the stellar response to the tidal forcing. These outputs can be used to compute secular evolution rates, including the circularization and synchronization timescales, which can be estimated via

\begin{equation}
\frac{1}{t_\mathrm{sync}} = \left| \frac{1}{\Omega_\mathrm{rot} - \Omega_\mathrm{orb}} \left[ \left(\frac{\mathrm{d}\Omega_\mathrm{rot}}{\mathrm{d} t}\right)_\mathrm{sec} - \left(\frac{\mathrm{d}\Omega_\mathrm{orb}}{\mathrm{d} t}\right)_\mathrm{sec} \right] \right|.
\end{equation}

The secular terms, $(\mathrm{d} \Omega_{\rm rot}/\mathrm{d} t)_{\rm sec}$ and $(\mathrm{d} \Omega_{\rm orb}/\mathrm{d} t)_{\rm sec}$, represent the long-term, averaged evolution rates of the stellar spin frequency and orbital frequency, respectively, driven by tidal dissipation. Their explicit expressions are provided below:

\begin{equation}
  \left(\frac{\mathrm{d}\Omega_\mathrm{rot}}{\mathrm{d} t}\right)_\mathrm{sec} = \frac{\mathcal{T}_{\rm sec}}{I},
\end{equation}
and,
\begin{equation}
    \left( \frac{\mathrm{d} \Omega_\mathrm{orb}}{\mathrm{d} t} \right)_\mathrm{sec} = -\frac{3}{2}\sqrt{G(M+M_{\rm comp})}a^{-5/2}\left(\frac{\mathrm{d} a}{\mathrm{d} t}\right)_\mathrm{sec}.
\end{equation}
where $\mathcal{T}_{\rm sec}$ is the secular tidal torque integrated over all relevant $(\ell, m, k)$ components. $G$ is the gravitational constant. In particular, $\mathcal{T}_{\rm sec}$ can be evaluated following Equation (25) of \citet{Sun2023}, then the secular evolution of the stellar rotation is

\begin{align}
\left(\frac{\mathrm{d} \Omega_{\rm rot}}{\mathrm{d} t}\right)_\mathrm{sec} 
&= 4 \Omega_{\rm orb} \sqrt{\frac{G M^2 M_{\rm comp}^{2}}{M + M_{\rm comp}}} 
q \frac{a^{1/2}}{I} \nonumber \\
&\quad \times \sum_{\ell=0}^{\infty} \sum_{m=-\ell}^{\ell} \sum_{k=0}^{\infty}
\left(\frac{R}{a}\right)^{\ell+3}
\left(\frac{r_b}{R}\right)^{\ell+1} \nonumber \\
&\quad \times \kappa_{\ell,m,k} \, |F_{\ell,m,k}| \, \sin \gamma_{\ell,m,k} \, G^{(4)}_{\ell,m,k}.
\end{align}
where $r_b$ is the boundary radius (which does not have to equal $R$ while we adopt the assumption of $r_b = R$), and $\gamma_{\ell,m,k} \equiv \arg(F_{\ell,m,k})$ is the phase lag associated with the complex response function $F_{\ell,m,k}$. The dimensionless factor $\kappa_{\ell,m,k}$, which accounts for the time-averaged strength of each partial tidal potential, is defined in Equation~(53) of \citet{Willems2010}, and the coefficient $G^{(4)}_{\ell,m,k}$ is given by their Equation~(64), encoding the efficiency of angular momentum transfer associated with each tidal component.

Using Equation (54) of \citet{Willems2010,Valsecchi2013} for $(\mathrm{d}a/\mathrm{d}t)_\mathrm{sec}$, $t_\mathrm{sync}$ has the final expression

\begin{equation}
\begin{aligned}
\frac{1}{t_\mathrm{sync}} = & \left|\frac{1}{\Omega_\mathrm{rot}-\Omega_\mathrm{orb}}\left[
\left(\frac{\mathrm{d}\Omega_\mathrm{rot}}{\mathrm{d} t}\right)_\mathrm{sec}\right.\right.\\
&\left.\left.\quad+\frac{3}{2}\sqrt{G(M+M_{\rm comp})}\,a^{-5/2}\left(\frac{\mathrm{d} a}{\mathrm{d} t}\right)_\mathrm{sec}\right]\right|\\[0.5ex]
= &\left|\frac{4\Omega_\mathrm{orb}}{\Omega_\mathrm{rot}-\Omega_\mathrm{orb}}q\sum_{\ell=0}^{\infty}\sum_{m=-\ell}^{\ell}\sum_{k=0}^{\infty}\right.\\
&\quad\times\left(\frac{R}{a}\right)^{\ell+3}\left(\frac{r_\mathrm{b}}{R}\right)^{\ell+1}\kappa_{\ell,m,k}|F_{\ell,m,k}|\sin\gamma_{\ell,m,k}\\
&\quad\times\left.\left(\frac{GMM_{\rm comp}}{aI\Omega_\mathrm{orb}}G^{(4)}_{\ell,m,k}+\frac{3}{2}\Omega_\mathrm{orb}G^{(2)}_{\ell,m,k}\right)\right|.
\label{tsync_numerical}
\end{aligned}
\end{equation}
Here, $G^{(2)}_{\ell,m,k}$ is a dimensionless coefficient describing the efficiency of orbital energy transfer due to each partial tide, as defined in Equation~(56) of \citet{Willems2010}. These coefficients, along with $F_{\ell,m,k}$, are directly output by the \texttt{GYRE-tides} module, allowing the user to compute the secular orbital evolution rates, including both $t_{\rm sync}$ and $t_a$,

\begin{equation}
\begin{split}
\frac{1}{t_a} 
= &\;  \left|\frac{1}{a} \left( \frac{da}{dt} \right)_{\rm sec}  \right| \\
= &\bigg|\; \frac{8\pi}{P_{\rm orb}} q 
  \sum_{\ell=2}^{\infty} \sum_{m=-\ell}^{\ell} \sum_{k=0}^{\infty} 
  \left( \frac{R}{a} \right)^{\ell+3} \\
& \times \kappa_{\ell,m,k} \, |F_{\ell,m,k}| \, 
  \sin \gamma_{\ell,m,k} \, G^{(2)}_{\ell,m,k}(e)\bigg| .
\label{ta_numerical}
\end{split}
\end{equation}

\section{Results\label{sec:results}}
We simulate a range of binary systems to compare secular evolution timescales using the \texttt{MESA} and \texttt{GYRE-tides} codes. Table~\ref{tab:binary_models} summarizes the selected binary models presented in this work. For each system, we compute the tidal response of the primary star as well as the orbital timescales $t_{\rm sync}$ and $t_{\rm a}$, while treating the secondary as a point mass. Both fixed-structure and time-evolving stellar models are considered, and all tidal calculations are performed on the primary, which is the initially more massive star.

For the numerical solution, we introduce a small eccentricity $e=0.01$ when feeding the problem into \texttt{GYRE-tides}. This is necessary because the Hansen coefficients cannot be evaluated accurately at very large values of $k$ for nearly circular orbit, a limitation of the current method. 
In the present work, we compute the Hansen coefficients as a function of $k$ and adopt a maximum value $k_{\max}=25$, adding tidal components only until the Hansen coefficient become negligibly small. {Additionally, the tidal secular evolution coefficients (e.g., \citealt{Willems2010}, their Eq.~57) contain factors of $1/e$; this choice prevents the expressions from diverging while having a negligible impact on our calculations of the orbital decay and synchronization timescales.} Unless otherwise stated, the tidal response considered in this section corresponds to the dominant $\ell = m = 2$ component. Including additional $m$ components does not significantly change the overall secular timescales, but can introduce extra resonant features at specific orbital periods.

\begin{table*}[t]
\centering
\caption{Initial Conditions of Binary Models Explored in This Work}
\label{tab:binary_models}
\begin{tabular}{ccccccc}
\hline\hline
$M_1/M_\odot$ & $M_2/M_\odot$ & $P_{\rm orb}$/days & Initially Synchronized & Section &Comments \\
\hline
5  & 1.4 & 0.5-10 & No  & \ref{Fixed Stellar Structure} & Fixed Stellar Structure \\
5 & 1.4  & 10  & Yes  & \ref{evol:5,1.4,10} & Evolving Systems \\
10 & 5  & 5  & Yes  & \ref{evol:10,5,5} &Evolving Systems \\
10 & 5  & 25  & Yes  & \ref{evol:10,5,25to50} & Evolving Systems \\
20 & 10  & 10  & Yes  & \ref{evol:20to50,10,10} & Evolving Systems \\
50 & 10  & 10  & Yes  & \ref{evol:20to50,10,10} & Evolving Systems \\
\hline
\end{tabular}
\medskip
\begin{flushleft}
\footnotesize\textit{Note:} We focus on tidal torques exerted on the primary $M_1$, while $M_2$ is treated as a point mass.
\end{flushleft}
\end{table*}

\subsection{Tidal Timescales for a Fixed Stellar Structure}
\label{Fixed Stellar Structure}

\begin{figure}[t]
    \centering
    \includegraphics[width=1\linewidth]{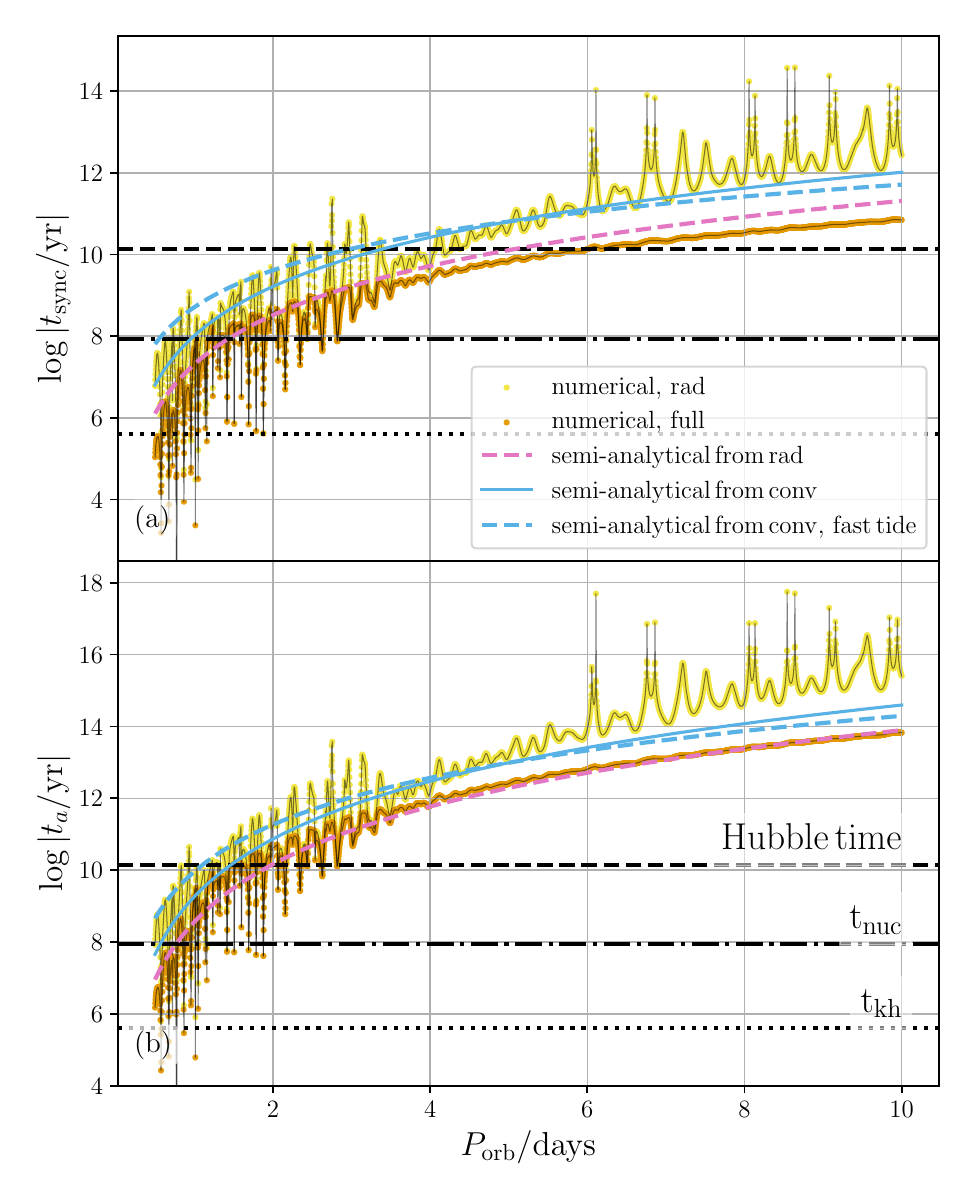}
    \caption{
Comparison of synchronization timescales (\textit{top}) and orbital decay timescales (\textit{bottom}) as functions of orbital period for a binary system with $M_1 = 5\,M_\odot$ and $M_2 = 1.4\,M_\odot$, evaluated at the ZAMS where $R_1 = 2.49\,R_\odot$, effective temperature $T_{\rm eff,1}= 17971\, {\rm K}$. The stellar rotation is assumed negligible initially, and the secondary is treated as a point mass. Only the orbital period is varied from 0.5 to 10 days. 
Results from the full direct solution (\texttt{GYRE-tides}; orange) are compared with those from the semi-analytical prescriptions for convective damping (blue) and radiative damping (pink). The yellow curves correspond to numerical results including only radiative diffusion damping, while the blue dashed lines indicate the semi-analytical prescription with the fast-tide treatment for convective damping. The Hubble time, thermal timescale $t_{\rm kh}$, and nuclear timescale $t_{\rm nuc}$ are shown as black dashed, dot-dashed, and dotted horizontal lines, respectively.
}
\label{fig:M1_5_M2_1.4_P_2to10_secular}
\end{figure}

To illustrate the impact of different tidal models, we calculate $t_{\rm sync}$ and $t_a$ for a $5\,M_\odot$ zero-age main-sequence (ZAMS) primary ($R_1 = 2.49\,R_\odot$ and $T_{\rm eff,1}= 17971\, {\rm K}$) with a $1.4\,M_\odot$ companion. The system is assumed to start with negligible stellar rotation, a circular and aligned orbit, and the secondary as a point mass. Only the orbital period is varied between 0.5 and 10 days. A similar mass configuration, representing a typical B-type star and a neutron star binary, was also considered in \citet{Sun2023} with a different eccentricity.

Figure~\ref{fig:M1_5_M2_1.4_P_2to10_secular} compares the synchronization ($t_\mathrm{sync}$, panel (a)) and semimajor axis evolution ($t_a$, panel (b)) timescales predicted by the \texttt{GYRE-tides} direct numerical solution (orange: full solution; yellow: radiative damping only) with those from the 
semi-analytical prescriptions for convective damping (blue) and radiative damping (pink). Across the range of $P_{\rm orb}\simeq 0.5$–10 days, the full numerical results (\texttt{GYRE-tides}; orange) are broadly consistent with the semi-analytical prescriptions. For synchronization timescales, the full numerical solution (including both damping mechanisms) agrees with the semi-analytical radiative damping prescription to within a factor of a few, with the numerical values remaining slightly shorter. The difference relative to the semi-analytical convective damping prescription is also modest, staying within an order of magnitude. For the semimajor-axis evolution timescale, the full numerical results and the semi-analytical radiative damping prescription are nearly indistinguishable, and differ from the semi-analytical convective damping case by less than one order of magnitude.

The numerical approach resolves resonances, producing spikes at 
$P_{\rm orb} \lesssim 3$--4 days that arise from internal gravity waves damped by radiative diffusion, whereas the semi-analytical prescription evaluated at a fixed stellar structure cannot capture such resonant features. It also delineates where the secular timescale computed with radiative damping alone (yellow) approaches the full solution including both radiative and convective damping (orange). At longer periods, the numerical curves become smooth, reflecting the dominance of the equilibrium tide, with dissipation set by the surface convective envelope. 

For a given stellar structure, the $(k/T)$ term can be fixed by radiative diffusion damping, so the synchronization timescale varies as a steep function of $P_{\rm orb}$ 
(specifically, $t_{\rm sync}\propto P_{\rm orb}^4$, and $t_a\propto P_{\rm orb}^{16/3}$ for the orbital evolution timescale). In the fast-tide limit, the convective contribution introduces an explicit $(k/T) \propto P_{\rm orb}$ dependence, giving $t_{\rm sync}\propto P_{\rm orb}^{3}$ and $t_a\propto P_{\rm orb}^{13/3}$. For this fixed $5\,M_\odot$ model, calibrated to a 5-day orbital configuration, we overplot a blue dashed line 
illustrating this scaling, in addition to the non–fast-tide prescription.


{While the semi-analytical prescription yields timescales that are numerically close to the numerical solution, this agreement mainly arises from a numerical similarity rather than the physical mechanism. In practice, the semi-analytical model always adopts the shorter of the radiative and convective contributions, so the radiative diffusion damping dominates at long forcing periods. In contrast, our numerical solution naturally includes both damping mechanisms and shows that, beyond \(P_{\rm orb}\!\gtrsim\!6\)~days, the convective damping contribution can become comparable to or even exceed the radiative one. This arises because the star retains a thin outer convection zone that, despite its small mass fraction, provides a non-negligible turbulent viscosity. The numerical tide may also involve the tail of the dynamical tide interacting with turbulent eddies in the evanescent zone, further enhancing dissipation in the outer layers. Therefore, the similarity between the semi-analytical and numerical results at longer periods should not be over-interpreted; it likely reflects the relative weighting of damping mechanisms rather than a physical equivalence between the two approaches.}

Physically, $t_a$ is generally longer than that for $t_{\rm sync}$, as has long been recognized from the ratio of orbital to spin angular momentum (see, e.g., Equation~35 in \citealt{Burkart2012}). In the present circular case, there is no distinct ``circularization timescale," but the qualitative separation of timescales persists: changes in the stellar spin rate always precede any substantial orbital evolution, owing to the much larger reservoir of angular momentum stored in the orbit relative to the star.

In summary, for this fixed $5\,M_\odot$ main-sequence model, the two approaches yield similar synchronization 
and orbital decay timescales overall, although the underlying damping mechanisms differ in their relative importance 
depending on $P_{\rm orb}$. Since the absolute values of these timescales often exceed the Hubble time or the stellar 
main-sequence (nuclear) lifetime (horizontal dashed lines), the remaining differences, within a factor of $\lesssim$10, have limited impact on binary evolution and population-synthesis studies. However, when the orbital period falls below about one day, the tidal secular evolution timescale can become comparable to the nuclear timescale or even the thermal timescale (dotted black line) with strong resonances, in which case tidal effects need to be considered even during the main-sequence phase.

\subsection{Comparing Tidal Timescales in Evolving Systems}

We first outline the method used to extract $t_{\rm sync}$ and $t_{\rm a}$ and the simulation setup. The binary orbital evolution is computed with \texttt{MESA}, which couples detailed stellar models from the binary module with the tidal prescriptions described in Section \ref{subsec:method:POSYDON-MESA}. Along the resulting evolutionary tracks, we then post-process each timestep with \texttt{GYRE-tides} to compute the tidal timescales defined in Section \ref{subsec:method:gyre}.

We first outline the method used to extract $t_{\rm sync}$ and $t_{\rm a}$ and the simulation setup. The binary orbital evolution {and spin evolution are} computed with \texttt{MESA}, which couples detailed stellar models from the binary module with the tidal prescriptions described in Section~\ref{subsec:method:POSYDON-MESA}. Along the resulting evolutionary tracks, we then post-process each timestep with \texttt{GYRE-tides} to compute the tidal timescales defined in Section~\ref{subsec:method:gyre}. 

This approach is motivated in part by the fact that \texttt{GYRE-tides} captures sharp resonances in the tidal response, which can lead to secular evolution timescales much shorter than the stellar evolution timescale at specific moments. Incorporating such resonant behavior into a fully coupled binary evolution scheme can result in numerical non-convergence and significantly increased computational cost. {As a result, the binary evolutionary trajectories shown in Figure~\ref{fig:M1_5_M2_1.4_P_10_secular} and subsequent figures are computed using the semi-analytical method. In principle, if these tidal numerical timescales were updated self-consistently to evolve the system, the orbital and spin periods could follow different trajectories at earlier epochs, causing the two methods to diverge.}

In the \texttt{MESA} binary tracks, systems are initialized with synchronized stellar spins and circular orbits. As stellar evolution progresses, when stellar radii increase and rotation slows, the system may deviate from synchronization. Tidal torques then act to restore synchronization, typically leading to orbital shrinkage.


\subsubsection{System Setup: $5\,M_\odot + 1.4\,M_\odot$, $P_{\rm orb} = 10$ Days}
\label{evol:5,1.4,10}

\begin{figure}[t]
    \centering
    \includegraphics[width=1\linewidth]{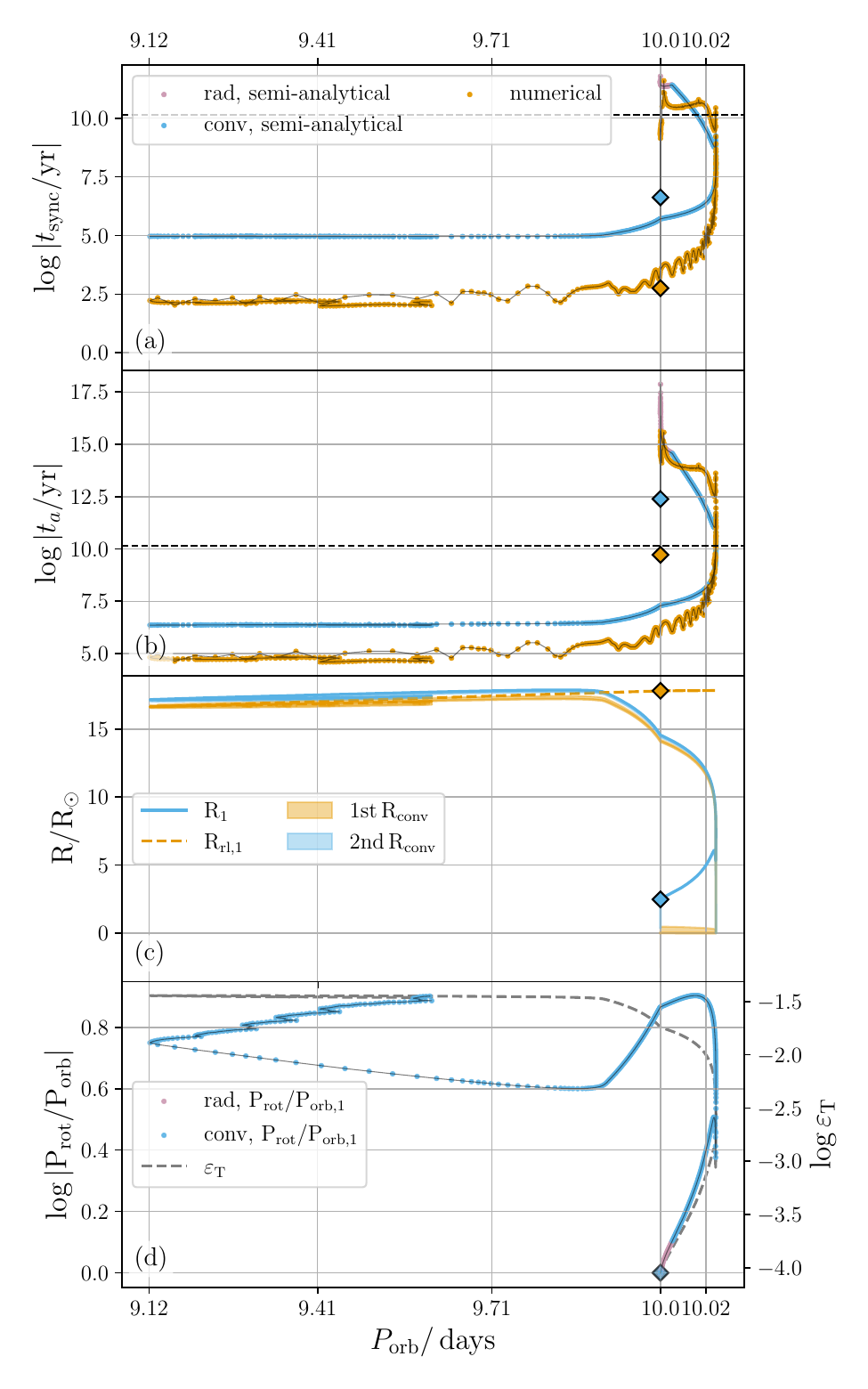}
    \caption{Tidal evolution diagnostics for a binary system with a $5\,M_\odot$ primary and a $1.4\,M_\odot$ companion, starting at an initial orbital period of 10 days. All panels are plotted against the orbital period $P_{\rm orb}$. Panels (a) and (b): Synchronization timescale $t_{\rm sync}$ and semimajor axis change timescale $t_a$ (both in years), showing \texttt{MESA} radiative damping dominated models (pink), \texttt{MESA} convective damping dominated models (blue), and \texttt{GYRE-tides} results (orange). The Hubble time is indicated by horizontal black lines. Panel (c): Stellar radius $R_1$ (blue), Roche-lobe radius $R_{\rm rl,1}$ (orange dashed), and convective zones (shaded), all in units of $R_\odot$. The orange shaded region corresponds to the most massive convective zone in radial coordinates (1st $R_{\rm conv}$), while the blue shaded region denotes the second most massive convective zone (2nd $R_{\rm conv}$). (d): Ratio of the primary spin period to the orbital period, $P_{\rm rot}/P_{\rm orb}$. The colored points show the evolution computed with radiative (pink) and convective (blue) damping using the semianalytical method. The tidal forcing strength $\varepsilon_\mathrm{T}$ is plotted as a gray dashed line. Diamond symbols mark the initial model.}
\label{fig:M1_5_M2_1.4_P_10_secular}
\end{figure}

We consider a binary consisting of a $5\,M_\odot$ primary and a $1.4\,M_\odot$ companion with an initial orbital period of 10 days. The secondary is treated as a point mass, and the orbit is assumed to be circular and aligned at the beginning of the evolution.

In Figure~\ref{fig:M1_5_M2_1.4_P_10_secular}, the horizontal axis represents $P_{\rm orb}$, which evolves over time. The top two panels compare $t_\mathrm{sync}$ and $t_a$ predicted by the numerical direct solution (orange) with those from the semi-analytical prescriptions for radiative damping (pink) and convective damping (blue). The results show that the numerical and semi-analytical methods are differ by about two orders of magnitude during the pre-RLOF phase, for both $t_\mathrm{sync}$ and $t_a$. However, during this stage, as the orbit first expands due to winds and later begins to shrink, the predicted timescales generally exceed the Hubble time. In such cases tides are not relevant to the binary evolution and differences between the two methods are inconsequential. Notably, the implied damping mechanisms differ: our results indicate that for $P_{\rm orb}\gtrsim5$ days the main sequence star responds in the long-wavelength regime, with dissipation dominated by equilibrium tides through convective damping even within the donor’s main-sequence lifetime, consistent with our previous findings in \citet{Sun2018}.

The numerical approach also resolves resonance features after the system reaches its maximum $P_{\rm orb}$ and the star enters the giant phase. These appear as modest bumps in $t_\mathrm{sync}$ and $t_a$ at $P_{\rm orb}\simeq 10.02$–10 days, corresponding to internal gravity waves damped by radiative diffusion. By contrast, the semi-analytical model cannot capture such resonances, and its predicted timescales vary smoothly as a function of orbital separation.

Panel (c) of Figure~\ref{fig:M1_5_M2_1.4_P_10_secular} shows the stellar radius $R_1$, the Roche-lobe radius $R_{\rm rl,1}$, and the two most massive convective zones in radial coordinates. The outer convective envelope remains thin throughout the main sequence. During the RLOF phase, both $t_\mathrm{sync}$ and $t_a$ from the numerical calculations are shorter than their semi-analytical counterparts by 1-2.5 orders of magnitude. In particular, $t_\mathrm{sync}$ reaches the order of $\sim10^3$ yr, suggesting that assuming strong tides are able to synchronize the system during RLOF may be reasonable. 

Panel (d) of Figure~\ref{fig:M1_5_M2_1.4_P_10_secular} shows that the system remains below synchronization ($P_{\rm rot}>P_{\rm orb}$) throughout the evolution, while the tidal forcing strength $\varepsilon_\mathrm{T}$ steadily increases as the orbit shrinks and the star expands. Along this evolutionary track, $\varepsilon_\mathrm{T}$ peaks at $\sim$0.03. The precise threshold for nonlinear effects is uncertain, values of $\varepsilon_\mathrm{T} \gtrsim 0.1$ are often considered indicative of the breakdown of linear tidal theory. For example, \citet{Ogilvie2014} discusses $\varepsilon_\mathrm{T} \sim 6 \times 10^{-2}$ in the context of hot Jupiter systems such as WASP-19 \citep{Hebb2010}. Our checking in $\varepsilon_\mathrm{T}$ suggesting that linear tidal theory remains marginally valid but may begin to underestimate dissipation during the most compact phases. {If the system were evolved with the numerically calculated torques, the faster orbital decay (smaller $a$) would increase $\epsilon_T \propto (R/a)^3$ 
and could lead to larger tidal amplitudes.}

Both panels (a) and (b) of Figure \ref{fig:M1_5_M2_1.4_P_10_secular} show an artificial jump at the first data point. On the numerical side, when the stellar spin and orbital frequencies are nearly equal, the result becomes numerically unstable (approaching a $0/0$ form). As a result, our \texttt{GYRE-tides} calculations can display noisy or spuriously small $t_{\rm sync}$ values when $\Omega_{\rm orb}\simeq\Omega_{\rm rot}$. Physically, however, the tidal timescale in this limit should be long, so these fluctuations reflect numerical sensitivity rather than true tidal torques. On the semi-analytical side, the initial jump originates from how $t_{\rm sync}$ is chosen: we ignore the core and select the convective layer that yields the shortest convective-damping timescale. In the very first model, the algorithm identifies a layer just outside the core, whose long turnover time relative to the surface layer leads to an artificially short $t_{\rm sync}$. From the next step onward, the initially selected convective layer near the core is absorbed into the growing core, and the calculation then identifies the surface convection zone instead. The timescales subsequently settle to the longer, physically relevant values seen thereafter. Consequently, the first timestep should not be used to interpret the secular evolution.

\begin{figure}[t]
    \centering
    \includegraphics[width=1\linewidth]{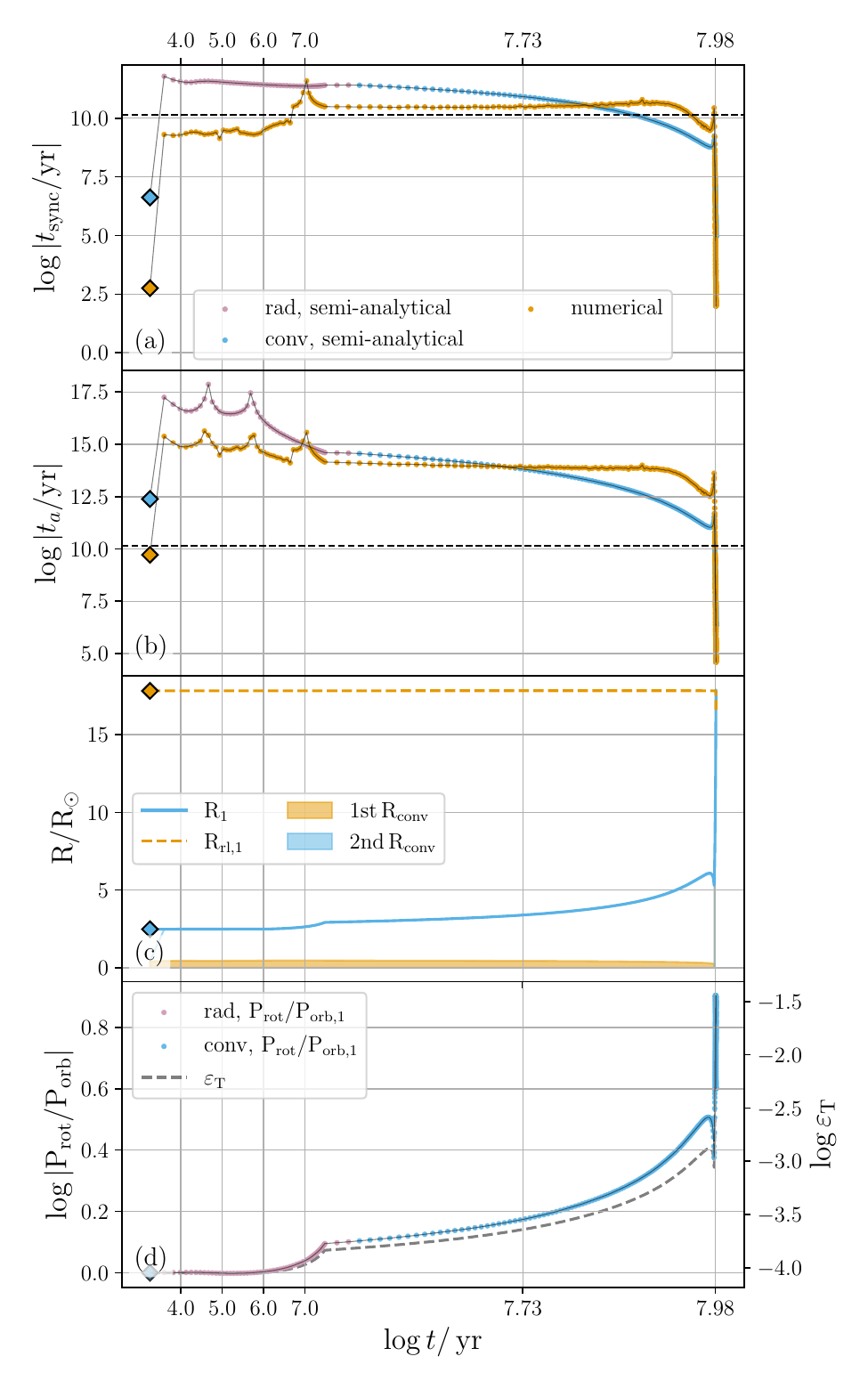}
    \caption{Same as Figure~\ref{fig:M1_5_M2_1.4_P_10_secular}, but with evolutionary time (in $\log\,t$/yr) on the horizontal axis instead of orbital period.}
\label{fig:M1_5_M2_1.4_P_10_t_secular}
\end{figure}

Figure~\ref{fig:M1_5_M2_1.4_P_10_t_secular} presents the same diagnostics as Figure~\ref{fig:M1_5_M2_1.4_P_10_secular}, but plotted against evolutionary time. This format highlights how the tidal timescales change relative to the internal structural evolution of the star before the onset of RLOF.

Panel (a) and panel (b) of Figure~\ref{fig:M1_5_M2_1.4_P_10_t_secular} show that before 10 Myr, both $t_\mathrm{sync}$ and $t_a$ differ by about two orders of magnitude, with the numerical solution implying stronger tidal dissipation. Several fluctuations in $t_a$ occur when $P_{\rm rot}\sim P_{\rm orb}$ for both the semi-analytical and numerical methods, or when resonances are excited. After 10 Myr and before RLOF, the discrepancies narrow. The semi-analytical method shows a steeper decline during the late main-sequence evolution of the $5\,M_\odot$ star, as also reflected in the increase of the stellar radius shown in panel (c) of Figure~\ref{fig:M1_5_M2_1.4_P_10_t_secular}. This behavior arises because the semi-analytical prescriptions impose a steeper $(R/a)$ dependence for $t_\mathrm{sync}$ and $t_a$, whereas in the numerical approach the dominant $\ell=2$ response, described by Equations~\ref{tsync_numerical} and \ref{ta_numerical}, scales approximately as $(R/a)^5$. After about 53 Myr, the numerical method even predicts a slower secular evolution rate.

Panel (c) of Figure~\ref{fig:M1_5_M2_1.4_P_10_t_secular} shows that the stellar radius $R_1$ gradually approaches the Roche-lobe radius $R_{\rm rl,1}$ as the primary evolves. The convective zones remain limited in mass until the late main-sequence phase, when only a thin surface convective envelope develops. Panel (d) of Figure~\ref{fig:M1_5_M2_1.4_P_10_t_secular} shows that the system remains persistently sub-synchronous ($P_{\rm rot}>P_{\rm orb}$). The tidal forcing strength $\varepsilon_\mathrm{T}$ increases as the orbit shrinks and the stellar radius grows, but remains well below 0.003, indicating that linear tidal theory may remain valid.

Overall, although the two approaches differ by up to about two orders of magnitude, before RLOF the timescales under this binary configuration are longer than the Hubble time. Thus, the remaining differences are probably negligible for population-synthesis studies and detailed binary evolution. During RLOF, however, the numerical results indicate that the system is likely to be more strongly synchronized. This implies that most systems may 
reach a quasi-steady orbital state earlier, consistent with the common assumption of strong tides during RLOF in binary evolution calculations. Therefore, these differences are unlikely to alter the broader conclusions drawn from population-synthesis models.

\subsubsection{System Setup: $10\,M_\odot + 5\,M_\odot$, $P_{\rm orb} = 5$ Days — Exploring an Intermediate-Mass Binary with a Close Orbit}
\label{evol:10,5,5}

Here we investigate a binary system composed of a $10\,M_\odot$ primary and a $5\,M_\odot$ compact object with an initial orbital period of 5 days. Compared to the previous $5\,M_\odot + 1.4\,M_\odot$ system with an initial orbital period of 10 days, such an intermediate-mass configuration is expected to experience stronger tidal interactions for two main reasons: (1) both objects are more massive, and the primary has a larger radius during much of the evolution, which increases the tidal torque; and (2) the closer initial separation enhances $\varepsilon_\mathrm{T}$ from the outset. The combination of these factors suggests that tidal synchronization and orbital decay should proceed more rapidly.

\begin{figure}[t]
    \centering
    \includegraphics[width=1\linewidth]{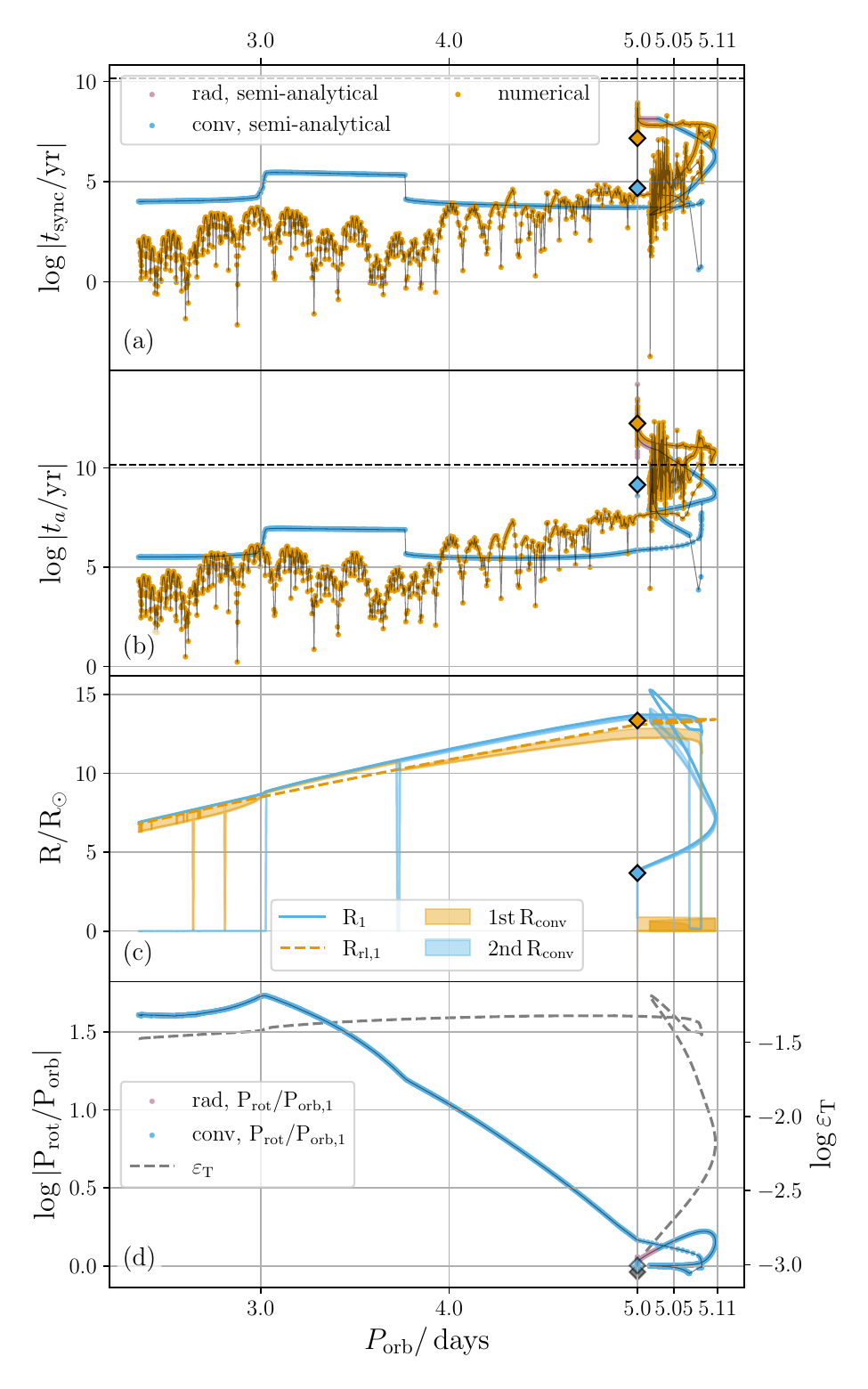}
    \caption{
Tidal secular evolution for a binary system with a $10\,M_\odot$ primary and a $5\,M_\odot$ companion, starting at an initial orbital period of 5 days. The format is the same as in Figure~\ref{fig:M1_5_M2_1.4_P_10_secular}.}
    \label{fig:M1_10_M2_5_P_5_secular}
\end{figure}

In this model, $P_{\rm orb}$ does not evolve monotonically but instead exhibits several turning points. For example, $P_{\rm orb}$ initially increases from 5.00 to 5.11 days as a result of strong wind-driven mass loss from the massive stars. This phase is followed by orbital shrinking, leading into the first RLOF episode near $P_{\rm orb} \simeq 5.04$ days. As shown in panel (c) of Figure \ref{fig:M1_10_M2_5_P_5_secular} with a clearer view, this is accompanied by a rapid expansion in the stellar radius $R_1$ and reach the Roche-lobe radius $R_{\rm rl,1}$. The orbit continues to shrink after mass transfer begins, reaching a minimum at $P_{\rm orb} \simeq 5.03$ days, after which it briefly re-expands before the first mass transfer phase ceases. A second phase of mass transfer starts around $P_{\rm orb} \simeq 5.09$ days, after which the system undergoes sustained orbital contraction until the end of the simulation.

During the brief expansion phase between 5.03 and 5.09 days, the primary spin period becomes shorter than the orbital period ($P_{\rm rot} < P_{\rm orb}$), as shown in panel (d). This temporary super-synchronous state is caused by a reduction in the stellar radius, from roughly $15\,R_\odot$ to $12.5\,R_\odot$, which increases $P_{\rm rot}$. In this interval, the semi analytical prescription predicts $t_{\rm sync} \sim 10^4$ years, short enough that strong tidal effects contribute to orbital expansion.

\begin{figure}[t]
    \centering
    \includegraphics[width=1\linewidth]{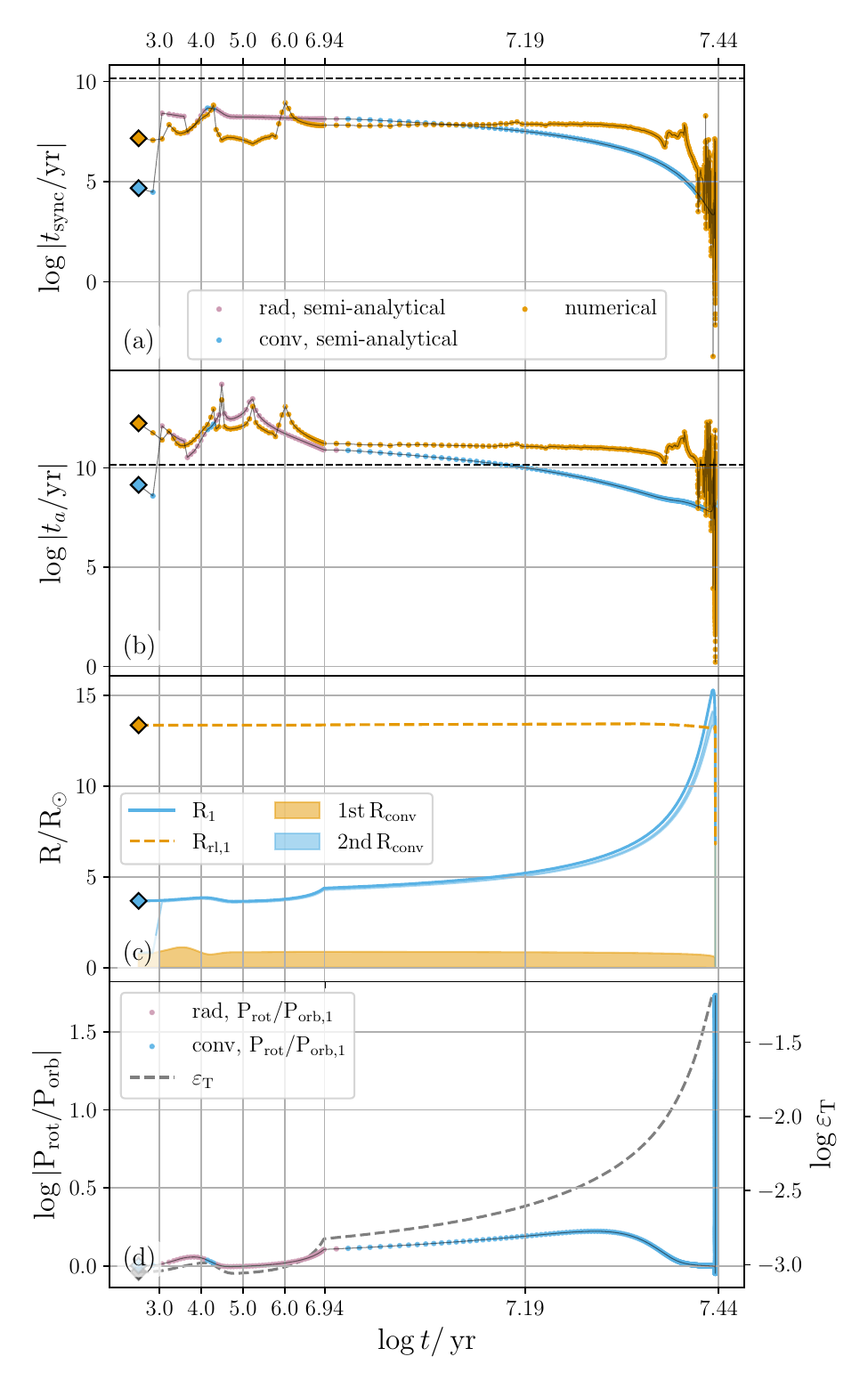}
    \caption{Same as Figure~\ref{fig:M1_10_M2_5_P_5_secular}, but with evolutionary time (in $\log\,t$/yr) on the horizontal axis instead of orbital period.
    } \label{fig:M1_10_M2_5_P_5_t_secular}
\end{figure}

Before RLOF, the top two panels of Figure~\ref{fig:M1_10_M2_5_P_5_secular} show that the numerical results predict tidal timescales similar to those from the semi-analytical model. Although the trends are not very clear in Figure~\ref{fig:M1_10_M2_5_P_5_secular}, readers can refer to Figure~\ref{fig:M1_10_M2_5_P_5_t_secular} for a clearer view. After RLOF, in the range of $P_{\rm orb}\simeq4$–5 days, the numerical $t_\mathrm{sync}$ and $t_a$ converge toward the semi-analytical predictions. When $P_{\rm orb}<4$ days, the numerical $t_\mathrm{sync}$ and $t_a$ are on average shorter by about 1–3 orders of magnitude. In addition, resonance-like fluctuations appear throughout the track, associated with internal gravity wave resonances.

Panel (c) of Figure~\ref{fig:M1_10_M2_5_P_5_secular} also shows the locations of the most massive and the second most massive convective zones, highlighted by the orange and blue shaded regions, respectively. Following the second phase of mass transfer, the surface convection zone becomes the largest convective region in the star. According to semi-analytical tides prescription, convective damping dominates the tidal dissipation during this stage. 

From panel (d) of Figure~\ref{fig:M1_10_M2_5_P_5_secular}, $\varepsilon_\mathrm{T}$ reaches a peak value of approximately 0.08 during the first phase of mass transfer. Prior to mass transfer, $\varepsilon_\mathrm{T}$ remains relatively low, indicating that the assumptions underlying linear tidal theory may remain valid in that regime.

Figure~\ref{fig:M1_10_M2_5_P_5_t_secular} presents the same system as Figure~\ref{fig:M1_10_M2_5_P_5_secular}, but plotted against evolutionary time. This format highlights how the tidal timescales evolve relative to the structural changes of the primary star before mass transfer. Panels (a) and (b) of Figure~\ref{fig:M1_10_M2_5_P_5_t_secular} show that most of the time during main sequence evolution, both $t_{\rm sync}$ and $t_a$ predicted by the numerical and semi-analytical models remain longer than the stellar main-sequence lifetime. This implies that tidal dissipation has a negligible effect on secular orbital evolution during the early main-sequence phase. As the star expands and approaches RLOF, both timescales drop substantially. The two methods remain broadly consistent in magnitude (typically within two orders), but differ in slope: the semi-analytical model shows a steeper decrease, while the numerical solution exhibits resonance-driven fluctuations.

Panel (c) of Figure~\ref{fig:M1_10_M2_5_P_5_t_secular} illustrates the internal convective structure. Before RLOF, the primary hosts two major convective zones: the most massive one located in the core, and a secondary zone near the stellar envelope.

Panel (d) of Figure~\ref{fig:M1_10_M2_5_P_5_t_secular} provides a clearer view of the stage leading up to the RLOF phase. The stellar rotation is updated according to the analytical $t_{\rm sync}$, which becomes short before RLOF, so the primary is effectively synchronized. When the stellar spin and orbital frequencies are nearly equal, the formal definition
$|t_{\rm sync}| = |(\Omega_{\rm orb}-\Omega_{\rm rot})/(\dot{\Omega}_{\rm orb}-\dot{\Omega}_{\rm rot})|$ becomes numerically unstable. As a result, our \texttt{GYRE-tides} calculations occasionally show noisy or artificially small values of $t_{\rm sync}$ when $\Omega_{\rm orb} \simeq \Omega_{\rm rot}$. Physically, the synchronization timescale in this regime should approach infinity, so these fluctuations mainly reflect numerical sensitivity rather than true tidal torques.

Finally, for the other binary configurations considered, we examined the same $10+5\,M_\odot$ system with wider initial orbits ($P_{\rm orb}=25$). We also explored the higher-mass $20+10\,M_\odot$ and $50+10\,M_\odot$ binaries. The qualitative behavior in these models is similar to the cases discussed above. Before RLOF, the numerical and semi-analytical approaches differ by about two orders of magnitude, but in these phases the absolute values of $t_{\rm sync}$ and $t_a$ remain longer than Hubble time, and the numerical solution can even yield slower secular evolution than the semi-analytical model. After the onset of RLOF, however, the numerical results consistently show a quite different orbital evolution rate than the semi-analytical prescription. A full set of diagnostics and figures for these configurations are provided in the Appendix, where the same trends can be examined in detail.

\section{Discussion\label{sec:discussions}}

\subsection{PSR J0045-7319 as a Benchmark: Validating the numerical method}
\label{subsec:PSR_J0045}


\begin{figure*}
    \centering
    \includegraphics[width=1\linewidth]{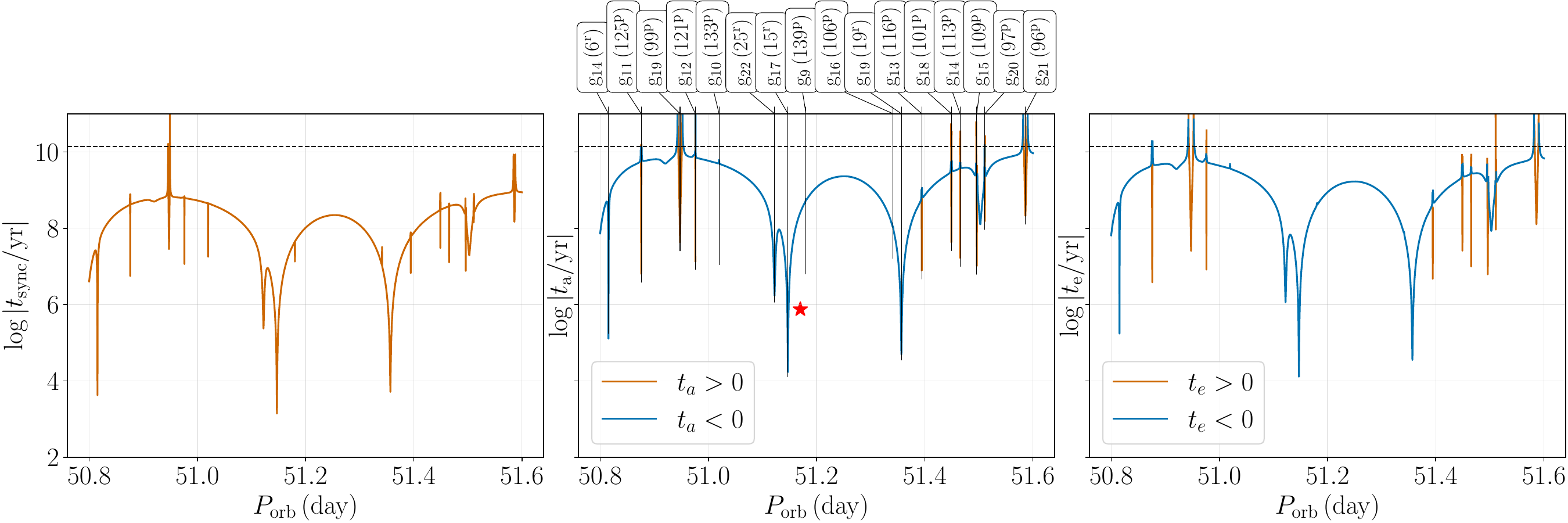}
    \caption{Tidal secular timescales for an $8.8\,M_\odot$ model with $R = 6.4\,R_\odot$, computed using \texttt{GYRE-tides}. Left panel: synchronization timescale $t_{\rm sync}$. Middle panel: semimajor axis evolution timescale $t_a$, with the red star indicating the observed orbital decay rate of PSR~J0045--7319 ($P_{\rm orb}/\dot{P}_{\rm orb} \simeq -0.5$ Myr). Right panel: circularization timescale $t_e$. Several prominent resonances are labeled in the upper region of the middle panel (see the text for a discussion of the labeling scheme.)}
    \label{fig:fig_secular_PSR_sys}
\end{figure*}

The PSR~J0045--7319 binary system provides a valuable observational benchmark for testing tidal dissipation models. First discovered by \citet{Kaspi1994}, it consists of a $1.4\,M_\odot$ radio pulsar and an $8.8\,M_\odot$ B-type companion in a highly eccentric ($e = 0.808$) orbit with a well measured orbital period of 51.17 days and a detected orbital decay rate of $P_{\rm orb}/\dot{P}_{\rm orb} \simeq -0.5$ Myr. The high eccentricity implies a much higher instantaneous angular velocity at periastron, equivalent to a local timescale of $P_{\rm peri} = (1-e)^{3/2}P_{\rm orb} = 4.3$ days. The stellar parameters of the B star are also well constrained: luminosity $L = 1.2 \times 10^4\,L_\odot$, $T_{\rm eff} = 24000 \pm 1000\,{\rm K}$, radius $R \approx 6.4\,R_\odot$, and projected surface rotation velocity $v\sin i = 113 \pm 10$ km/s \citep{Bell1995}, consistent with moderately rapid rotation.

This system has already been well studied by \citet{Lai1996, Lai1997} using a modal decomposition method. Additional work by \citet{Kumar1997, Kumar1998} included damping calculations, showing that radiative damping alone is insufficient to explain the observed orbital decay rate. Nonlinear damping mechanisms—such as the traveling wave approximation or parametric instability—as well as significant differential rotation are likely required. \citet{Su2022} presented an analytical expression for the tidal torque exerted on stars by dynamical tides in eccentric massive binaries. To explain the PSR~J0045--7319 system within that framework, the B star must exhibit substantial retrograde and differential rotation, or possess a significantly large convective core.

To explore whether our models can account for the observed orbital decay in PSR~J0045--7319, we constructed a representative \texttt{MESA} model (version 24.08.1) by evolving an $8.8\,M_\odot$ zero-age main sequence star until its radius reached $6.4\,R_\odot$, and then analyzed its tidal response using \texttt{GYRE-tides}. The \texttt{MESA} and \texttt{GYRE} inlists used in this section are publicly available on Zenodo with the DOI \href{https://doi.org/10.5281/zenodo.17941341}
{10.5281/zenodo.17941341}. We adopted $P_{\rm rot} = 1.73$ days, assuming an inclination angle of $44^\circ$ (co-planar with the orbital plane), which implies a super-synchronous rotation: that is, the stellar spin frequency exceeds the instantaneous orbital angular frequency at periastron. Figure~\ref{fig:fig_secular_PSR_sys} presents the predicted synchronization, semimajor axis evolution, and circularization timescales as functions of orbital period near the observed value. These results are obtained by fixing the stellar structure and varying the orbital period input to the tidal evolution calculation, and under the assumption that the $\ell = 2$ components of the tidal potential dominate the tidal coupling. The red star marks the observed orbital decay timescale of $P_{\rm orb}/\dot{P}_{\rm orb} \simeq -0.5$ Myr.

In our definition, $t_a > 0$ and $t_e > 0$ indicate orbital decay and decreasing eccentricity, respectively (see \citealt{Sun2023}, Equation~22, for details). Considering the absolute value of $t_a$, we find that while the upper envelope of $t_a$ values predicted by \texttt{GYRE-tides} still differs from the observed orbital decay timescale by approximately three orders of magnitude, several localized minima, corresponding to near-resonant interactions with stellar oscillation modes, approach the measured value. These results highlight the requirement for exceptionally resonant g-modes to enable efficient tidal dissipation in such systems. One notable example is the $g_{17}(15\mathrm{r})$ resonance --- in the nomenclature adopted by \citet{Sun2023}, arising when the the $k=15$ harmonic of the orbital frequency drives the g-mode with radial order $n=17$ that has a retrograde propagation direction in the co-rotating frame. Most of the spikes originate from resonances with very high-order orbital harmonics ($k \sim 100$). {We note that, given the secular spin down of the star and the expansion of the orbit, the forcing frequencies sweep across the g-mode spectrum rather than co-evolve with individual modes, and the intrinsic g-mode frequencies in massive main-sequence stars tend to increase slowly as the Brunt--V\"ais\"al\"a frequency rises near the contracting convective core \citep{Wu2018}. As a result, sustained resonance locking is unlikely in this system, and the observed spikes are more naturally interpreted as brief resonance crossings.}

In this study, our focus is on interpreting the absolute value of the orbital decay rate. 
{For the PSR~J0045--7319 system, the B-type primary rotates faster than the orbital angular frequency at periastron. 
In a simple equilibrium-tide picture, the tidal torque on the star is negative: angular momentum is extracted from the stellar spin and transferred to the orbit, causing the orbit to expand and the eccentricity to increase. 
However, in our GYRE-tides calculations, the resonant dynamical tides can still remove orbital energy and orbital angular momentum (corresponding to the orange lines in the middle and right panel of Figure \ref{fig:fig_secular_PSR_sys}), so that the net effect is an orbital decay. 
In such cases, all of the relevant resonances involve prograde and overstable modes. 
This behaviour is qualitatively similar to the ``inverse-tide'' regime discussed by \citet{Fuller2021}, and warrants further investigation.}

The results shown in Figure~\ref{fig:fig_secular_PSR_sys} were obtained using a purely radiative damping prescription. We find that including convective damping yields nearly identical results, confirming that the dominant tidal dissipation mechanism is radiative damping of internal gravity waves.

This study represents an initial exploration of the tidal dynamics in this system; future work will aim to incorporate additional physical effects and achieve a more quantitative match to observations. Our current calculations neglect the Coriolis force, which can significantly alter the frequencies of gravity modes and may also affect their damping rates in rotating stars. However, incorporating the Coriolis force within the direct solution approach of \texttt{GYRE-tides} is not straightforward. This is because modifying the surface boundary condition of the Poisson equation to include the companion's tidal potential is incompatible when using Hough functions, which does not permit separation of variables in the Poisson equation \citep{Fellay2025}. Moreover, although we include the stellar rotation rate as an input parameter to compute the tidal forcing frequency, we assume rigid-body rotation and do not account for differential rotation or possible evolution in obliquity. 

It is also important to note that the stellar evolution models used in this work were generated using the rotation mixing prescription adopted in the \texttt{MIST} framework, which is implemented in \texttt{POSYDON-MESA} tracks. While this prescription performs reasonably well for general stellar evolution calculations, it is not optimized for asteroseismic applications. In particular, the resulting Brunt–Väisälä frequency profiles often exhibit unphysical spikes, which can distort the eigenmode structure and lead to inaccurate identification of resonant modes. As a result, caution is warranted when interpreting the mode identifications in our current analysis. Future work should revisit these calculations with improved internal structure models better suited for tidally driven wave diagnostics.

Nevertheless, our results provide compelling support for the physical plausibility of resonance-driven orbital decay in massive binaries and highlight the benefits of employing a fully resolved, frequency-dependent tidal formalism. Semi-analytical tidal prescription tends to yield orbital decay timescales exceeding $10^{14}$ years, far longer than typical stellar lifetimes. These findings underscore the importance of incorporating more realistic tidal dissipation models into future binary population synthesis efforts.

Similar to how the PSR~J0045--7319 system serves as a benchmark for high-mass binaries, open cluster circularization periods provide a crucial constraint on tidal efficiency in low-mass populations. In open clusters, where stars are coeval and have similar metallicities, the observed distribution of eccentricity versus $P_{\rm orb}$ reveals a characteristic circularization period (CP), a threshold below which most binaries are circularized. This provides a population-level constraint on tidal dissipation.

\citet{Geller2013} and \citet{Milliman2014} showed that the binary population synthesis code \texttt{BSE} significantly underpredicts the observed CP for main-sequence binaries. To match the observations, they had to artificially increase the tidal damping efficiency by a factor of 100. This suggests that, during the main-sequence phase, current tidal prescriptions in population synthesis frameworks may be too weak to reproduce the observed CPs \citep{Meibom2005}.

\subsection{Possible Resonance Locking}

In some of our simulations, we observe sharp drops in both $t_{\rm sync}$ and $t_a$. {However, we caution that these features should not be interpreted as true resonance locks. A self-consistent treatment of resonance locking (or resonance crossing) would require that the evolutionary timescales $t_{\rm sync}$ and $t_a$ computed directly with the numerical method be incorporated into an integration of the orbital evolution, to determine whether a sustained lock is maintained.} Resonance locking has been systematically explored in a variety of astrophysical contexts, including heartbeat stars, compact object binaries, exoplanetary systems, and gas giant moon systems \citep{Burkart2012, Burkart2014, Fuller2012, Fuller2016, Fuller2017, Ma2021, Zanazzi2021}. In these cases, the stellar eigenfrequency and the tidal forcing frequency evolve in the same direction (e.g., both increasing), allowing the system to remain near resonance over secular timescales. This process enables sustained angular momentum transfer between the orbit and the star by maintaining the tidal forcing near a resonant mode frequency, thereby significantly amplifying the efficiency of tidal dissipation \citep{Witte1999, Burkart2012, Fuller2012, Fuller2017}.

In our models, these features arise from the direct solution approach of \texttt{GYRE-tides}, which captures the full frequency-dependent tidal response and allows for resonant behavior to emerge naturally. In contrast, parameterized prescriptions cannot resolve individual resonances and thus fail to reproduce such rapid tidal evolution.

During the early evolution of massive binaries, before mass transfer, the primary star expands and loses mass via stellar winds, leading to a decrease in its dynamical frequency $\sqrt{GM/R^3}$ and, consequently, a downward shift in its mode frequencies. Meanwhile, both the orbital frequency $\Omega_{\rm orb}$ and the stellar spin frequency $\Omega_{\rm rot}$ typically decrease, causing the tidal forcing frequency $\sigma_{m,k} = k\Omega_{\rm orb} - m\Omega_{\rm rot}$ to decrease as well in common cases where $k = m$. This is only a qualitative expectation: the relative rates at which $\Omega_{\rm orb}$ and $\Omega_{\rm rot}$ evolve also matter, and contributions from other harmonics may dominate the dissipation. In addition, during the post-main-sequence evolution of massive stars, compositional gradients tend to develop due to the lack of efficient mixing between nuclear burning shells. These gradients lead to an increase in the Brunt–Väisälä frequency, which in turn can raise the frequencies of gravity modes, as g-mode oscillations are primarily sensitive to the buoyancy profile. 

Whether resonance locking can be maintained for a given mode depends on detailed quantities such as the overlap integrals and damping rates. At this stage, we have not yet carried out mode identification to determine which oscillations are involved. A detailed asteroseismic analysis of the mode spectrum and coupling strengths is deferred to future work. In the absence of resonance locking, systems are expected to undergo resonance crossing, where the forcing frequency briefly sweeps through mode frequencies without locking; see \citet{Ma2023, Ma2024}.

\subsection{Caveat on Nearly Circular Orbits and Hansen Coefficient Truncation}
For nearly circular orbits analyzed with the numerical approach, only a few low-$k$ Hansen coefficients contribute significantly to the tidal potential, since the Hansen coefficients decay rapidly with $k$. However, the numerical evaluation of Hansen coefficients at very small $e$ can become unstable for large $k$, even though the true coefficients should be vanishingly small. To avoid spurious contributions from this numerical noise, we precompute Hansen coefficients and truncate the sum at a $k_{\max}$ where the coefficients reach the level of code precision.

This is a limitation of the current numerical method, to varying degrees, by most techniques in the literature, which are typically optimized for small $k$. We are developing a more robust treatment to extend the reliable range in $k$ without resorting to artificially large $e$, and will report on this in future work.

\section{Conclusions\label{sec:conclusions}}
In this work, we systematically compare the tidal secular evolution timescales predicted by the direct numerical method implemented in \texttt{GYRE-tides} with the semi-analytic prescription adopted in the binary evolution code \texttt{MESA} also used the binary population synthesis code \texttt{POSYDON}. We examine across a range of configurations involving intermediate- to high-mass stars ($M_1 = 5$–$50\,M_\odot$). We focus on systems with initially circular orbits and synchronized rotation, and follow their evolution through mass loss, stellar expansion, and, in many cases, mass transfer. For each system, we extract stellar structure snapshots from \texttt{MESA} simulations and compute synchronization and semimajor-axis evolution timescales at every timestep using both methods. This allows us to directly quantify discrepancies between the two prescriptions and assess their potential impact on binary evolutionary outcomes. We further apply this comparative methodology—evaluating tidal evolution using both the direct numerical calculations from \texttt{GYRE-tides} and the semi-analytic prescription—to the well-characterized PSR~J0045--7319 system, using it as an observational benchmark to assess the reliability of both approaches.

Our main findings are as follows:

\begin{itemize}

\item \textbf{Pre–RLOF regime.} For both $t_{\rm sync}$ and $t_a$, the numerical and semi-analytical results differ by about 1--2 orders of magnitude during the main-sequence evolution prior to RLOF. In these phases, the absolute timescales are typically longer than the Hubble time (or at least the stellar nuclear lifetime), so these differences are inconsequential for binary evolution and population-synthesis studies.

\item \textbf{Resonant structure.} The direct solution naturally resolves resonance features (seen as modest excursions rather than sharp spikes in fixed structure scans), arising from g–mode excitation and radiative damping. Semi–analytical prescriptions cannot capture these mode effects and therefore produce smooth, monotonic scalings with $P_{\rm orb}$.

\item \textbf{During RLOF.} In the $10\,M_\odot$ and $5\,M_\odot$ donor models, once mass transfer begins, the numerical results predict $t_{\rm sync}$ and $t_a$ shorter by about 1--2.5 orders of magnitude compared to the semi-analytical estimates. For $t_a$, in particular, the differences can reach an order of magnitude. This indicates stronger synchronization during RLOF for these systems. This implies that most systems may reach a quasi-steady orbital state earlier, consistent with the common assumption of strong tides during RLOF in binary evolution calculations. Therefore, these differences are unlikely to alter the broader conclusions drawn from detailed binary modeling and population-synthesis models. Nevertheless, their ultimate importance depends on comparisons with other stellar timescales, such as the mass transfer timescale, global thermal timescale and the local thermal timescale in the tidal dissipation region, which we leave for future work.

\item \textbf{Dominant Damping Mechanisms May Differ.} Despite the magnitude level agreement, the implied physics is not the same. For $P_{\rm orb}\gtrsim5$\,d the star resides in the long wavelength regime and dissipation is dominated by equilibrium tides with convective damping over the donor’s main–sequence lifetime. At shorter periods ($\lesssim3$–4\,d), radiative damping of gravity waves becomes increasingly important.

In the semi-analytical prescription, the shortest synchronization timescale among the available convective 
layers is adopted as the global $t_{\rm sync}$, which effectively places the dominant dissipation in the thin surface convection zone. This treatment is sensitive to arbitrary criteria (e.g., the 10–cell threshold adopted in \texttt{POSYDON-MESA}). By contrast, the numerical calculations resolve the stellar oscillation spectrum and consistently find radiative damping of gravity waves to be the main dissipation channel in close massive binaries. This difference reflects the limitations of the parametrization, which compresses the complex tidal response into a single fitting function tied to global stellar properties rather than tracking where energy is actually deposited.

In reality, the physical location where tidal energy is converted and deposited remains uncertain, even for dynamical tides alone. In solar-like stars with radiative cores and convective envelopes, internal gravity waves excited near the RCB can propagate inward. As these waves steepen and grow in amplitude, nonlinear dissipation may cause them to deposit energy deep within the stellar interior. In intermediate- and high-mass stars with convective cores and radiative envelopes, outgoing internal gravity waves are subject to radiative damping as they propagate outward. Because the thermal timescale becomes shorter toward the stellar surface, high-order, long-period gravity modes are particularly vulnerable to damping in these outer layers. Depending on the wave properties and boundary conditions, wave energy may be absorbed near the surface or escape entirely from the star. In the latter case, total energy conservation is broken within the stellar system. Pinpointing the precise dissipation sites and quantifying their contributions are key areas for future investigation.

\item \textbf{Applicability of Linear Tidal Theory.} We evaluate the dimensionless tidal forcing strength $\varepsilon_{\rm T}$ in each system to assess the validity of linear theory. In many binaries, $\varepsilon_{\rm T}$ remains well below 0.01 throughout the evolution, indicating that linear tidal theory may remain applicable. However, in some compact configurations, particularly those involving $M_1 \gtrsim 20\,M_\odot$ and $P_{\rm orb} \lesssim 10$ days, $\varepsilon_{\rm T}$ can approach 0.1, suggesting that nonlinear effects may become important and that linear models may underestimate dissipation.

\item \textbf{Case Study of the PSR~J0045--7319 System.} This analysis demonstrates the potential of fully resolved tidal calculations to capture the complex behavior of dynamical tides. Although our treatment is preliminary and does not yet include Coriolis effects or refined stellar structure models, the results suggest that resonant interactions with internal gravity waves can play a decisive role in the secular evolution of massive binaries. In particular, \texttt{GYRE-tides} tentatively reproduces the observed orbital period change rate through strong resonances that are entirely missed by parameterized prescriptions. The discrepancy extends beyond resonance behavior: even in non-resonant regimes, the semi-analytical tidal prescription yields orbital change timescales exceeding $10^{14}$ years, highlighting its limitations. These findings underscore the predictive advantages of direct numerical methods in modeling specific systems. Future work incorporating more realistic stellar structures and rotation physics will be essential to establish a comprehensive picture of resonance-driven tidal dissipation in eccentric, massive binaries.

\end{itemize}

Overall, our results suggest a two–tier approach. For population–synthesis studies, where computational speed is essential, parameterized prescriptions remain practical and. A modest re-calibration may be needed when the predicted tidal timescales are either much longer than the Hubble time or shorter than the relevant stellar timescales. However, when investigating individual systems in detail, simplified formulas are inadequate: only direct numerical methods can reliably capture the resonant features and the underlying physics of tidal dissipation. In this sense, numerical approaches are indispensable for interpreting observations of specific binaries, while population level studies may continue to rely on improved parameterizations.

The discrepancies likely stem from the different ways in which tidal dissipation is approximated in the two methods. Although Zahn's original formulation was based on modal decomposition, many subsequent studies have fit $E_2$ as a function of stellar mass or radius using simplified assumptions. As highlighted by \citet{Kushnir2017}, ``such parameterizations can lead to inconsistencies and order-of-magnitude uncertainties in torque estimates, particularly when applied outside the regimes where the fits were calibrated". Our findings reinforce the importance of accounting for the internal structure of the star and resolving mode properties, especially in systems where tidal dissipation plays a key role in secular evolution.

In addition, part of the discrepancy arises from how each method handles dissipation. Semi-analytical method selects the shortest $t_{\rm sync}$, often from a surface convective layer, which may underestimate radiative damping deeper inside the star. By contrast, numerical method solves the full non-adiabatic response and integrates contributions from internal gravity modes across the interior, capturing frequency dependent effects. This often results in stronger torques and radiative damping dominance. We also note that widely used prescriptions, such as \citet{Hurley2002}, have inherited errors in implementing Zahn’s formalism. As pointed out by \citet{Sciarini2024}, these can misrepresent circularization timescales.

Zahn’s pioneering tidal framework dates to the 1960s, when limited computational resources necessitated semi-analytic prescriptions and parameterized fits derived under restrictive assumptions. With today’s open-source tools, we can evolve detailed binaries and resolve mode spectra even on personal machines, enabling a re-examination of those prescriptions across much broader regions of parameter space. Looking ahead, our comparison highlights that existing semi-analytical prescriptions often reproduce numerical timescales to within two orders of magnitude across much of the parameter space. The main distinction lies in the underlying damping mechanisms assumed to dominate in different regimes. As such, the priority is to clarify the regimes where different mechanisms (radiative vs.\ convective) provide the correct physical interpretation. Recent efforts in this direction include semi-analytic dissipation models calibrated to numerical results, such as those proposed by \citet{Kushnir2017} and \citet{Su2022}, and applications to extremely close systems by \citet{Ma2023}. Incorporating such prescriptions into population-synthesis frameworks still offer a practical compromise, capturing the relevant physics while remaining computationally efficient. A systematic comparison between \texttt{GYRE-tides}, classical analytic models, and these emerging intermediate prescriptions will be crucial for identifying which approaches are both robust in practice and physically well-motivated across the massive-binary parameter space.

\section*{Acknowledgments}

We thank the referee for the careful reading of the manuscript and for several insightful suggestions that helped improve the clarity and accuracy of the work. This work is supported by the National Natural Science Foundation of China through the Fundamental Science Center for nearby Galaxy (grant No. 12588202), a project based on LAMOST and FAST. M.S., H.X., S.G., and K.A.R. thank the support from the Gordon and Betty Moore Foundation (PI: Vicky Kalogera, grant number GBMF8477). M.S. extends appreciation to the entire \texttt{POSYDON} developer team for their invaluable technical assistance. This work was performed in part at Aspen Center for Physics, which is supported by National Science Foundation grant PHY-2210452, Simons Foundation (1161654, Troyer) and Alfred P. Sloan Foundation (G-2024-22395). S.G. acknowledges funding support from a CIERA Postdoctoral Fellowship. V.K. acknowledges partial support from the D. I. Linzer Distinguished University Professorship Fund. K.A.R. thanks the support by the CIERA’s Riedel Family Fellowship and the NASA grant awarded to the Illinois/NASA Space Grant Consortium. R.H.D.T. acknowledges support from NASA grants 80NSSC24K0895 and 80NSSC23K1517. E.Z. acknowledges support from the Hellenic Foundation for Research and Innovation (H.F.R.I.) under the “3rd Call for H.F.R.I. Research Projects to support Post-Doctoral Researchers” (Project No: 7933).

Special thanks are extended to Phil Arras, Janosz Dewberry, Dong Lai, Yoram Lithwick, Norm Murray, Gordon Ogilvie, Fred Rasio, Yubo Su and Yanqin Wu for their insightful comments and stimulating discussions on various aspects of stellar tides. Their expertise, from analytical tidal theory to numerical modeling and secular evolution—greatly informed the development of this work. We also thank Zhao Guo and Linhao Ma for helpful discussions on resonance locking. M.S. acknowledges Chenliang Huang for carefully checking key derivations. 

\software{\texttt{numpy} \citep{harris2020array}, \texttt{Matplotlib} \citep{Hunter2007}, \texttt{pandas} \citep{reback2020pandas}, 
\texttt{GYRE} \citep{Townsend2013,Townsend2018,Goldstein2020,Sun2023}, \texttt{MESA} 
\citep{Paxton2011,Paxton2013,Paxton2015,Paxton2018,Paxton2019,Jermyn2023}, \texttt{POSYDON} \citep{Fragos2023,Andrews2024}.}

\section*{Appendix A \label{sec:appendixA}}
\setcounter{figure}{0}
\renewcommand{\thefigure}{A\arabic{figure}}

\setcounter{subsection}{0}
\renewcommand{\thesubsection}{A\arabic{subsection}}

The results presented in this Appendix follow the same analysis as in Section~\ref{sec:results}. Overall, the conclusions remain qualitatively similar: the two tidal prescriptions predict similar synchronization and orbital-decay timescales, typically differing by within 1–3 orders of magnitude. Since these trends are consistent with the cases already discussed in the main text, we relegate the following examples to the Appendix. They are included here for completeness, in case readers are interested in inspecting the detailed differences for additional binary configurations.

\subsection{System Setup: $10\,M_\odot + 5\,M_\odot$, $P_{\rm orb} = 25$ Days — Wider Orbit}
\label{evol:10,5,25to50}

\begin{figure}[t]
    \centering
    \includegraphics[width=1\linewidth]{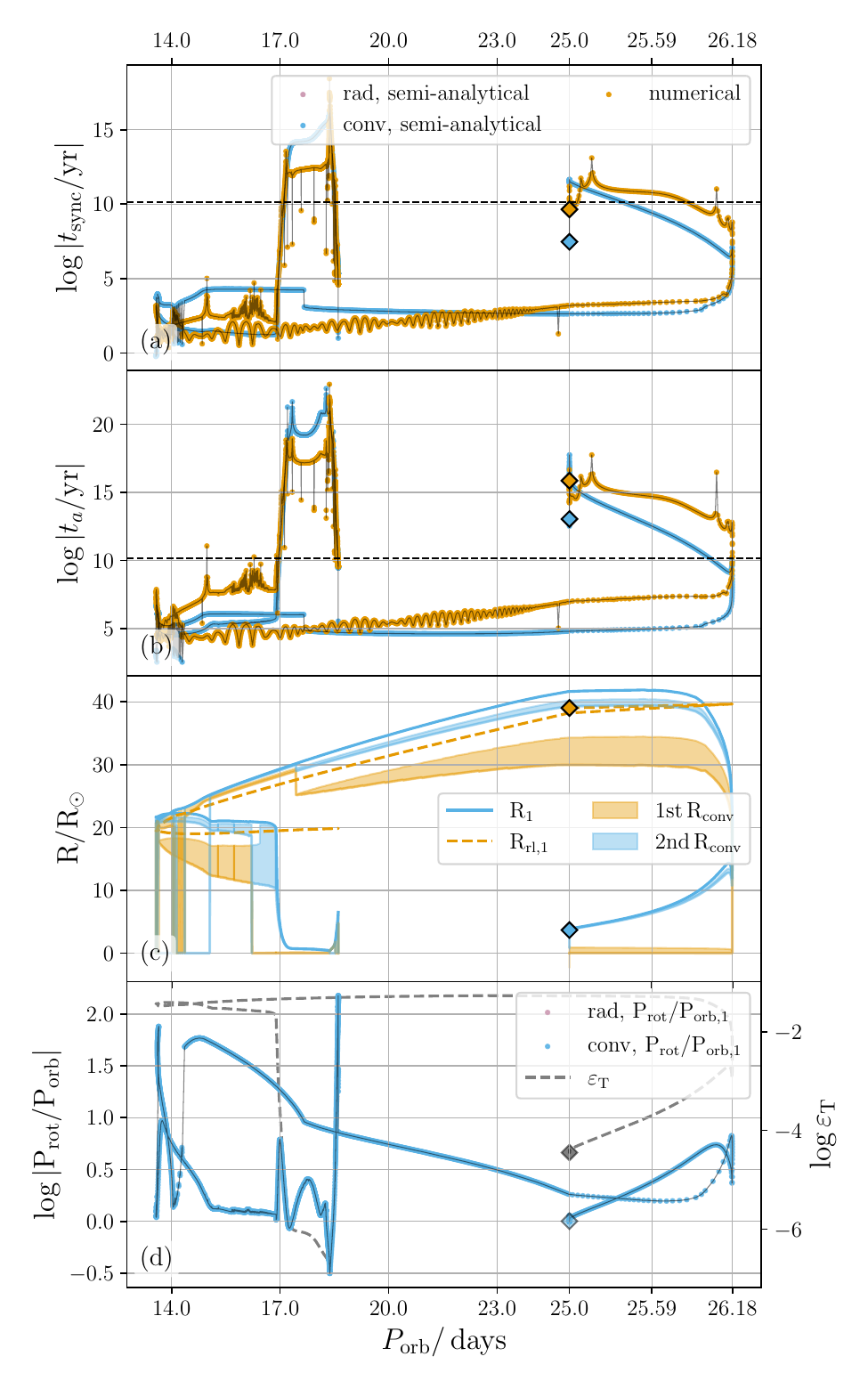}
    \caption{Tidal secular evolution for a binary system with a $10\,M_\odot$ primary and a $5\,M_\odot$ companion, starting at an initial orbital period of 25 days. The format is the same as in Figure~\ref{fig:M1_10_M2_5_P_5_secular}.}
    \label{fig:M1_10_M2_5_P_25_secular}
\end{figure}

We now explore the case of a wider binary orbit, starting at $P_{\rm orb} = 25$ days, while keeping the component masses fixed at $10\,M_\odot$ and $5\,M_\odot$. Compared to the tighter orbit of 5 days discussed in Section~\ref{evol:10,5,5}, tidal interactions in this configuration are substantially weaker due to the increased orbital separation. Initially, the orbital period increases slightly, reaching a maximum near $P_{\rm orb}\simeq26.2$ days before decreasing. RLOF occurs at approximately $P_{\rm orb}\simeq25.8$ days, as indicated by Panel (c), where the primary radius equals its Roche-lobe radius ($R_1 = R_{\rm rl,1}$), and also marked by the red ``+" symbol in panel (b). The orbit continues shrinking after the onset of mass transfer until reaching about $P_{\rm orb}\simeq14$ days, then re-expands slightly as mass transfer ceases around 17 days.

Figure~\ref{fig:M1_10_M2_5_P_25_secular} shows that $t_{\rm sync}$ and $t_a$ are significantly longer than in closer-orbit systems, especially at early times. In particular, both the semi-analytical and numerical methods predict $t_{\mathrm{sync}}$ values ranging from $\sim10^6$ to $10^{11}$ years prior to RLOF. For $t_a$, the discrepancy between the two approaches is comparable to that for $t_{\rm sync}$, typically spanning about 1–3 orders of magnitude. Before RLOF, the two methods give $t_a$ values longer than the Hubble time, indicating negligible tidal influence on orbital evolution.

During the RLOF phase, the discrepancies in $t_{\mathrm{sync}}$ and $t_a$ between the two models are on the order of 1–3 magnitude. After mass transfer ceases, however, the differences increase to approximately 2–3 orders of magnitude, but this is not significant because the corresponding timescales are far longer than the age of the Universe. Panel (c) shows that, when plotted against $P_{\rm orb}$, the star possesses two convective zones for most of the evolution: the most massive one at the core and a secondary, less massive zone near the surface. Across most of the $P_{\rm orb}$ parameter space, the semi-analytical method attributes the dominant tidal dissipation to convective damping.

Panel (d) shows that the system remains persistently sub-synchronous throughout most of its evolution until mass transfer ends. Before RLOF, $\varepsilon_\mathrm{T}$ remains below $10^{-2}$, suggesting linear tidal theory is a suitable approximation. During mass transfer, however, $\varepsilon_\mathrm{T}$ increases to values between 0.01 and 0.1, indicating that tidal dissipation may be underestimated due to neglected nonlinear effects. After mass transfer ends, the primary star rapidly shrinks in radius, causing $P_{\rm rot}$ to decrease sharply, temporarily becoming shorter than $P_{\rm orb}$. At this stage, tidal effects become negligible, as both $t_{\mathrm{sync}}$ and $t_a$ exceed the age of the universe. Consequently, the observed orbital expansion following mass transfer cessation is not tidally driven.

\begin{figure}[t]
    \centering
    \includegraphics[width=1\linewidth]{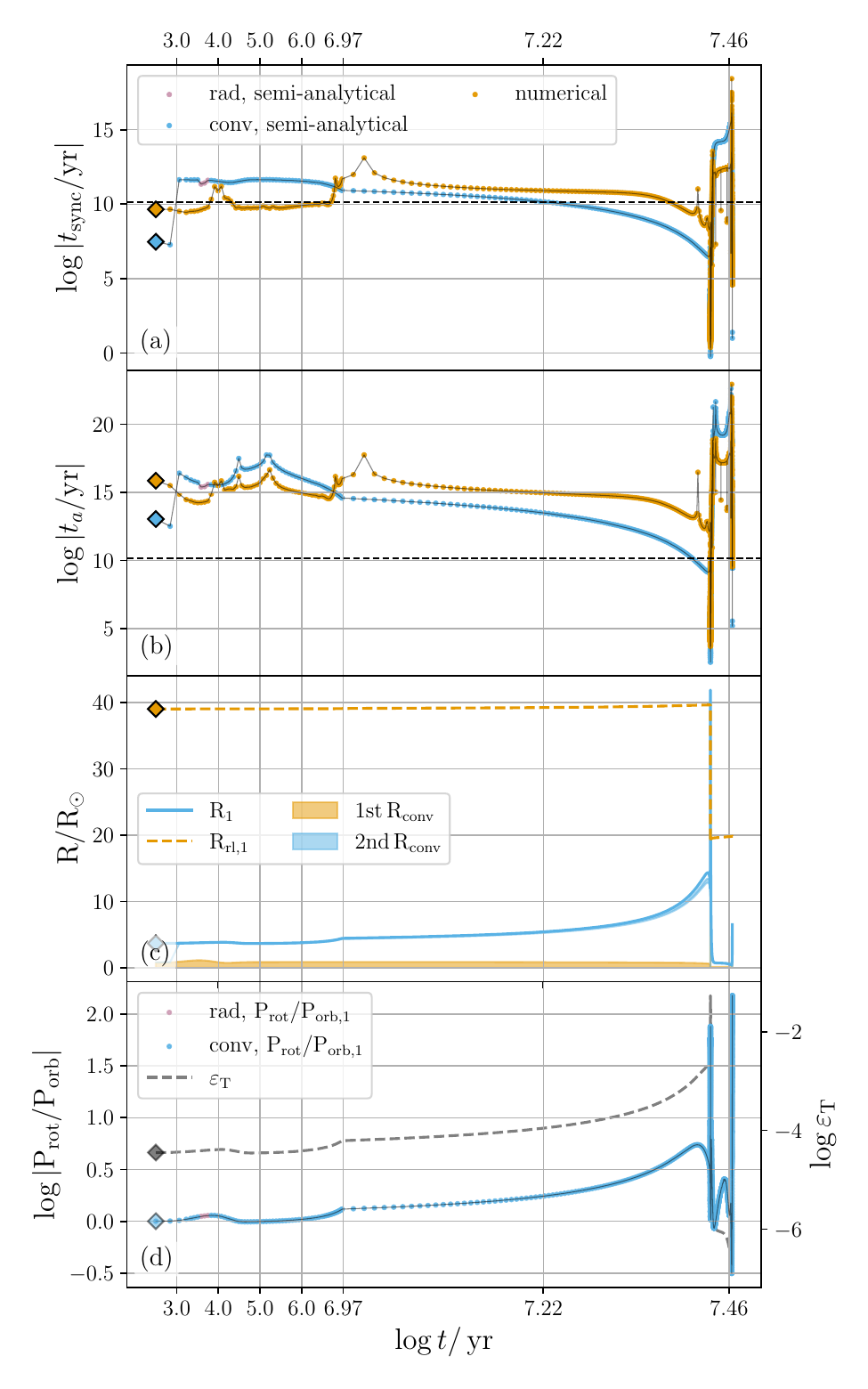}
    \caption{Same as Figure~\ref{fig:M1_10_M2_5_P_25_secular}, but with evolutionary time (in $\log\,t$/yr) on the horizontal axis instead of orbital period.
    } \label{fig:M1_10_M2_5_P_25_t_secular}
\end{figure}

Figure~\ref{fig:M1_10_M2_5_P_25_t_secular} displays the same tidal diagnostics plotted against evolutionary time. This time-based representation more clearly resolves the tidal evolution and resonance structure at earlier stages. Before roughly 10 Myr, the numerical method predicts smaller $t_{\rm sync}$ and $t_a$, differing by about two orders of magnitude. After that, the numerical results yield longer $t_{\rm sync}$ and $t_a$, but for most of the evolution both timescales remain longer than the age of the Universe. Therefore, for nearly circular orbits in this configuration, tidal effects play a negligible role in altering the orbital evolution.


\subsection{System Setup: $20\,M_\odot$ and $50\,M_\odot + 10\,M_\odot$, $P_{\rm orb} = 10$ — Massive Binary with a Close Orbit}
\label{evol:20to50,10,10}

We also examine an extremely massive binary with a $50\,M_\odot$ donor and a $10\,M_\odot$ companion, starting at an orbital period of $P_{\rm orb} = 10$ days. Due to the increased stellar masses, tidal interactions are stronger compared to the less massive systems with similar initial orbital periods discussed previously (e.g., the $5\,M_\odot$ primary with a $1.4\,M_\odot$ secondary). Because $\varepsilon_\mathrm{T}$ reaches values of order $\sim0.1$ during RLOF, the validity of linear tidal theory may be limited in scope. For this reason, we only show tidal timescales as a function of time, focusing on the pre-RLOF phase.

\begin{figure}[t]
    \centering
    \includegraphics[width=1\linewidth]{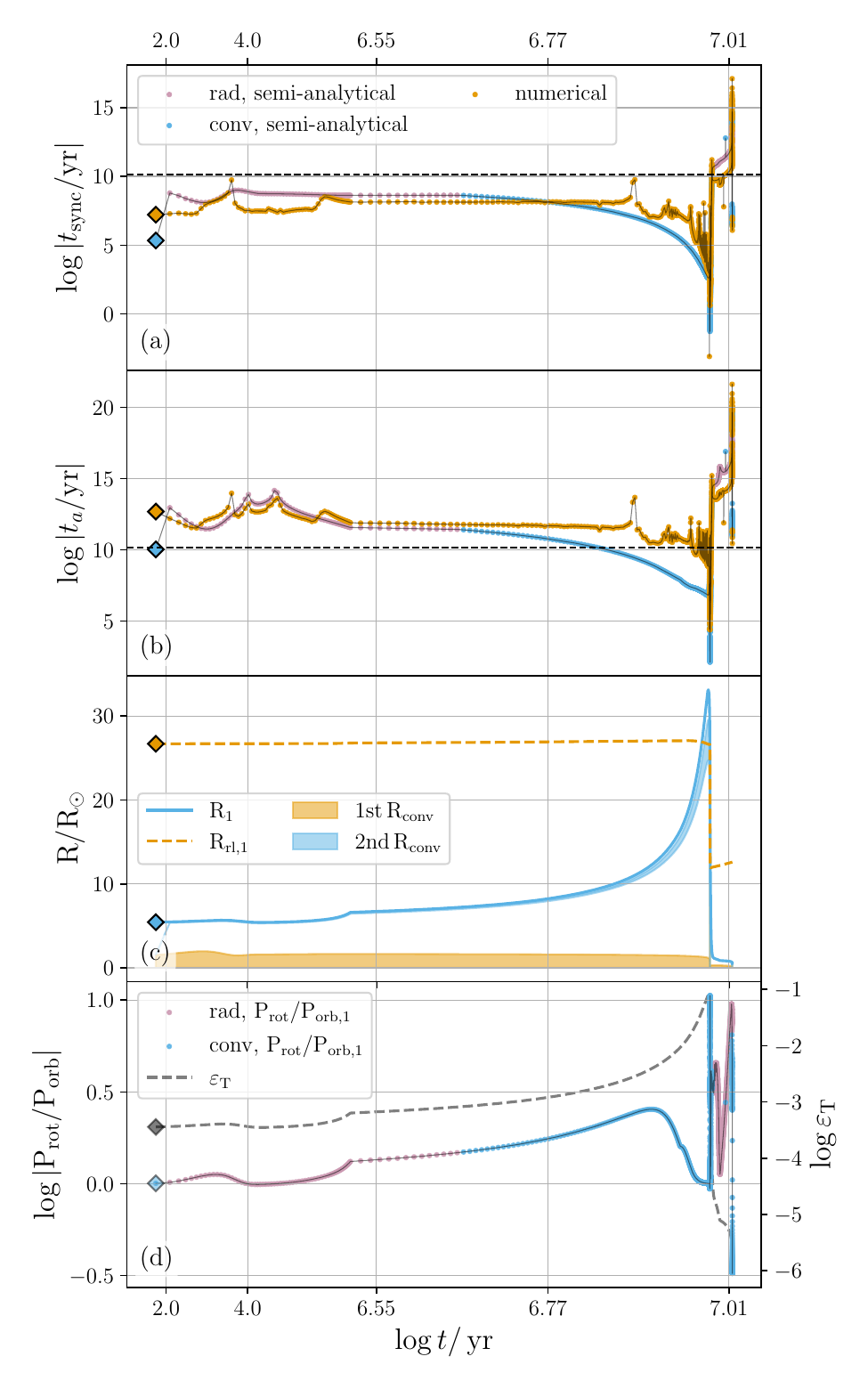}
    \caption{Tidal secular evolution for a binary system with a $20\,M_\odot$ primary and a $10\,M_\odot$ companion, starting at an initial orbital period of 10 days, with evolutionary time (in $\log\,t$/yr) on the horizontal axis.
} \label{fig:M1_20_M2_10_P_10_t_secular}
\end{figure}

Figure~\ref{fig:M1_20_M2_10_P_10_t_secular} presents the tidal secular evolution for a binary system with a $20\,M_\odot$ donor and a $10\,M_\odot$ companion at an initial orbital period of 10 days. Panels (a) and (b) compare $t_{\rm sync}$ and $t_a$, respectively, as predicted by the numerical method (orange) and by the semi-analytical prescriptions for radiative (pink) and convective (blue) damping. During most of the pre-RLOF evolution, the two approaches yield similar $t_{\rm sync}$ and $t_a$ values. Just before the onset of RLOF, however, the semi-analytical results drop more steeply, while the numerical solution shows resonance features. Since $t_a$ remains longer than the Hubble time for most of the track, tidal effects can generally be neglected when considering orbital evolution in this configuration.  

Panel (c) of Figure \ref{fig:M1_20_M2_10_P_10_t_secular} shows that the donor maintains two major convective zones: a massive convective core and a thinner convective envelope near the surface. In the semi-analytical framework, the surface convection zone dominates the dissipation, leading to systematically shorter synchronization and orbital-decay timescales than in the numerical results. Panel (d) shows that the system remains sub-synchronous ($P_{\rm rot}>P_{\rm orb}$) throughout most of the track. The tidal forcing strength $\varepsilon_{\rm T}$ steadily grows as the donor expands and the orbit contracts, reaching $\sim0.1$ near RLOF. This suggests that nonlinear effects may become important in this phase, so the linear analysis presented here may underestimate tidal dissipation once Roche-lobe overflow begins.

\begin{figure}[t]
    \centering
    \includegraphics[width=1\linewidth]{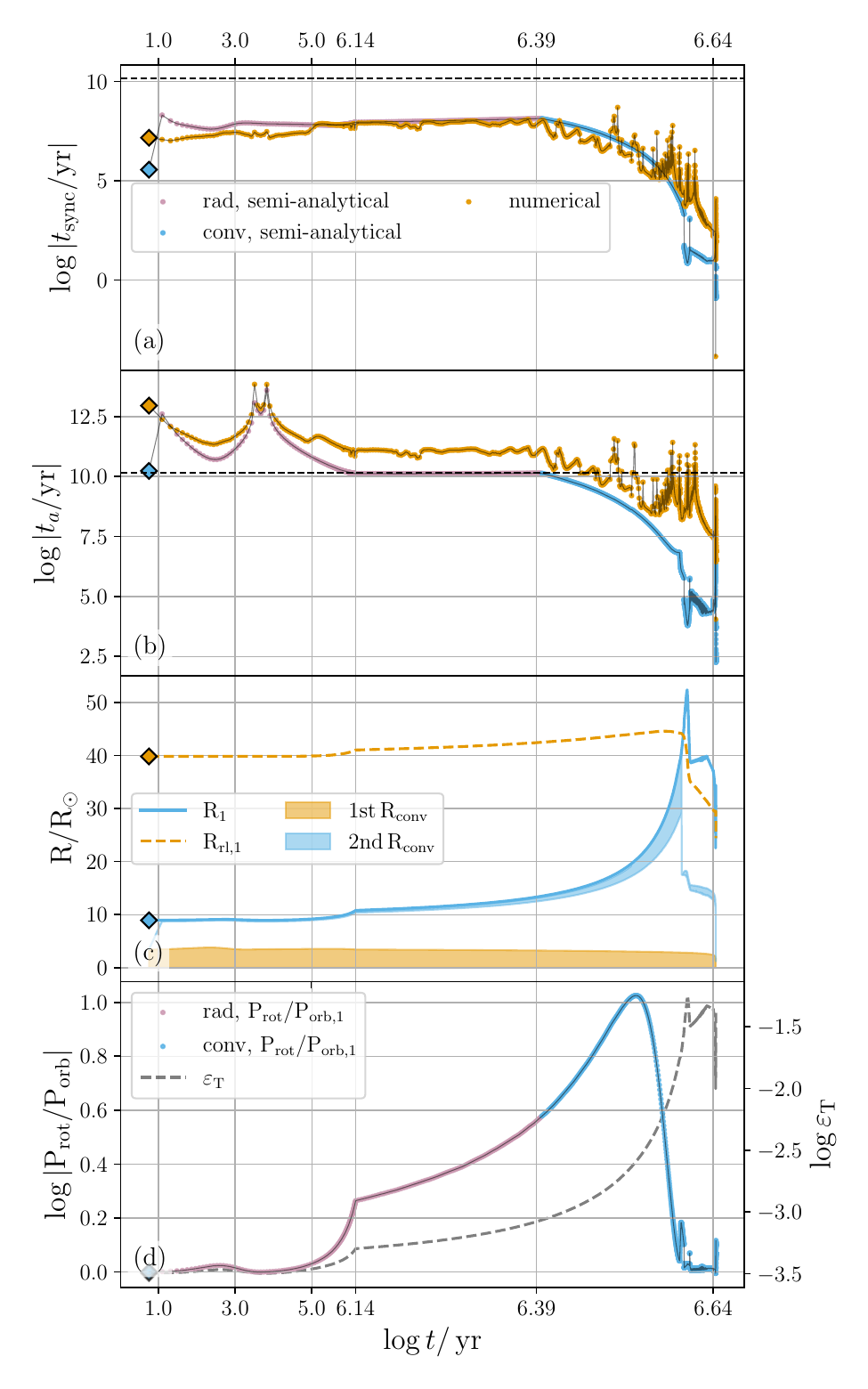}
    \caption{Tidal secular evolution for a binary system with a $50\,M_\odot$ primary and a $10\,M_\odot$ companion, starting at an initial orbital period of 10 days, with evolutionary time (in $\log\,t$/yr) on the horizontal axis.}
    \label{fig:M1_50_M2_10_P_10_t_secular}
\end{figure}

Figure~\ref{fig:M1_50_M2_10_P_10_t_secular} presents the tidal secular evolution for a binary system with a $50\,M_\odot$ donor and a $10\,M_\odot$ companion at an initial orbital period of 10 days. Panels (a) and (b) compare $t_{\rm sync}$ and $t_a$, respectively, from the numerical method (orange) with the semi-analytical prescriptions for radiative (pink) and convective (blue) damping. For most of the pre-RLOF evolution, the two approaches yield broadly similar values, though the numerical results display modest resonance features absent from the smooth semi-analytical curves near the onset of RLOF. Before $\sim$2.5 Myr, $t_a$ remains longer than the Hubble time, indicating that tidal dissipation is inefficient in shaping the orbital evolution. After $\sim$2.5 Myr, the two methods predict similar trends for $t_{\rm sync}$, both dropping rapidly to values below 0.1 Myr, implying efficient synchronization of the system, despite the resonance fluctuations in the numerical results. For $t_a$, however, the semi-analytical model predicts systematically shorter timescales, by about 1–3 orders of magnitude, compared to the numerical method.

Panel (c) of Figure~\ref{fig:M1_50_M2_10_P_10_t_secular} shows that the donor retains two convective zones: a massive convective core and a relatively thin convective envelope near the surface. In the semi-analytical framework, the surface convection zone dominates the dissipation after $\sim$2.5 Myr, leading to shorter $t_{\rm sync}$ and $t_a$ values. Panel (d) demonstrates that the system remains sub-synchronous ($P_{\rm rot}>P_{\rm orb}$) throughout most of the pre-RLOF phase. $\varepsilon_{\rm T}$ increases steadily as the star expands and the orbit shrinks, reaching values close to $\sim0.1$ near the onset of RLOF. This again suggests that nonlinear effects may become significant at this stage, and linear tidal theory could underestimate the true dissipation efficiency once mass transfer begins.

\bibliography{references.bib}{}
\bibliographystyle{aasjournal}

\end{CJK*}
\end{document}